\documentclass[fleqn,usenatbib]{mnras}

\usepackage{newtxtext,newtxmath}

\usepackage[T1]{fontenc}

\DeclareRobustCommand{\VAN}[3]{#2}
\let\VANthebibliography\thebibliography
\def\thebibliography{\DeclareRobustCommand{\VAN}[3]{##3}\VANthebibliography}

\usepackage{graphicx}	
\usepackage{amsmath}	
\newcommand*{\wasyfamily}{\fontencoding{U}\fontfamily{wasy}\selectfont}
\newcommand*{\mercury}{{\text{\wasyfamily\char39}}}

\title[NS with off-centre dipole magnetic field]{Three-dimensional magneto-thermal evolution of off-centred dipole magnetic field configurations in neutron stars}

\author[A.P. Igoshev et al.]{
Andrei P. Igoshev,$^{1,2}$\thanks{E-mail: a.igoshev@leeds.ac.uk, ignotur@gmail.com}
Rainer Hollerbach$^{1,2}$
\& Toby Wood$^{1,3}$
\\
$^{1}$Isaac Newton Institute for Mathematical Sciences, 20 Clarkson Road, Cambridge CB3 0EH, UK; \\
$^{2}$Department of Applied Mathematics, University of Leeds, Leeds LS2 9JT, UK\\
$^{3}$School of Mathematics, Statistics and Physics, University of Newcastle, Newcastle upon Tyne NE1 7RU, UK
}

\date{Accepted XXX. Received YYY; in original form ZZZ}

\pubyear{2015}

\begin{document}
\label{firstpage}
\pagerange{\pageref{firstpage}--\pageref{lastpage}}
\maketitle

\begin{abstract}
Off-centred dipole configurations
have been suggested 
to explain different phenomena in neutron stars, such as natal kicks,  irregularities in polarisation of radio pulsars and properties of X-ray emission from millisecond pulsars. Here for the first time we model magneto-thermal evolution of neutron stars with crust-confined magnetic fields and off-centred dipole moments.
We find that the dipole shift decays with time if the initial configuration has no toroidal magnetic field. The decay timescale is inversely proportional to magnetic field. The octupole moment decreases much faster than the quadrupole. Alternatively, if the initial condition includes strong dipolar toroidal magnetic field, the external poloidal magnetic field evolves from centred dipole to off-centred dipole. The surface thermal maps are very different for configurations with weak $B = 10^{13}$~G and strong $B = 10^{14}$~G magnetic fields. In the former case, the magnetic equator is cold while in the latter case it is hot. We model lightcurves and spectra of our magneto-thermal configurations. We found that in the case of cold equator, the pulsed fraction is small (below a few percent in most cases) and spectra are well described with a single blackbody. Under the same conditions models with stronger magnetic fields produce lightcurves with pulsed fraction of tens of percent. Their spectra are significantly better described with two blackbodies. Overall, the magnetic field strength has a more significant effect on bulk thermal emission of neutron stars than does the field geometry.  
\end{abstract}




\section{Introduction}

Magnetic fields largely define the observational properties of isolated neutron stars (NSs), including their thermal properties, radio pulsar activity and polarisation. Differences in magnetic field geometry 
and evolution 
are thought to
explain much of the 
observed diversity of isolated NSs \citep{Harding2013FrPhy}.
The initial magnetic field geometry and its evolution are still matters of active scientific research \citep[for a review see e.g.][]{Igoshev2021Univ}. 

It is typically assumed that magnetic fields of neutron stars are
predominantly dipolar. 
The dipole moment is well measured via the spin-down of isolated neutron stars, and is directly proportional to the square-root of measured period $P$ and period derivative $\dot P$: $B_p \propto \sqrt{P\dot P}$ \citep{OstrikerGunn1969ApJ,Philippov2014MNRAS}. The reason for this is that higher-order moments decay faster with distance from the NS and become weak at the light cylinder where pulsar braking takes place. 

\begin{figure}
    \centering
    \includegraphics[width=\columnwidth]{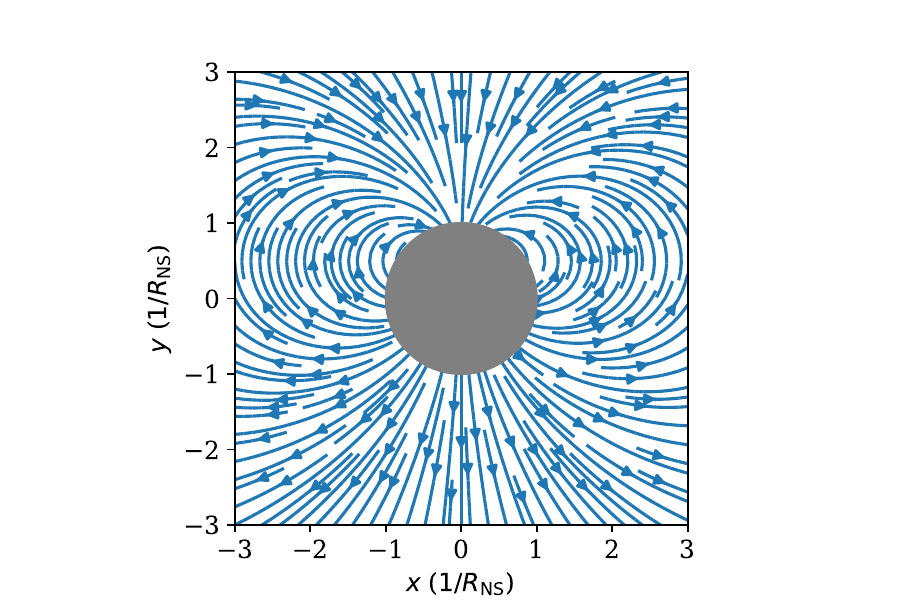}
    \caption{External field lines for an off-centred dipole, shifted in the polar direction by $d=0.5$~R$_\mathrm{NS}$. The grey circle shows the size of the neutron star.}
    \label{fig:field_lines}
\end{figure}

An off-centred dipole configuration was originally suggested by \cite{Harrison1975ApJ} to explain fast speeds of NSs, so-called natal kick \citep{Lyne1994Natur} thus formulating electromagnetic rocket scenario; modern estimates for natal kick distribution are provided by \cite{Verbunt2017AA,Igoshev2020MNRAS}. 
Asymmetries of electromagnetic emission could accelerate a neutron star over time, so that it reaches velocities comparable to hundreds of km/s. Very recently \protect\cite{AgalianouGourgouliatos2023arXiv} further investigated this scenario and found that the dipole shift should be comparable to $\approx 7$~km ($\approx 0.6$~$R_\mathrm{NS}$) from the rotational axis to explain typical NS velocities. 
The same idea was also 
used to explain irregularities in the behaviour of radio polarisation \citep{BurnettMelatos2014MNRAS}. Motivated by these ideas, \cite{Petri2016MNRAS}
derived an analytic expression for the electromagnetic radiation produced
by an off-centred dipole configuration. 
We show an example of the field lines for an off-centred dipole in vacuum in Figure~\ref{fig:field_lines}.

The Neutron star Interior Composition Explorer (NICER) X-ray observatory \citep{Gendreau2016SPIE} observed millisecond radio pulsar PSR J0030+0451. \cite{Miller2019ApJ,Riley2019ApJ} estimated the mass and radius of this pulsar using the phase-resolved spectroscopy and analytical modelling for photon propagation in curved spacetime. They noticed that two hot spots at the surface of this NS are not symmetric with respect to NS rotational axis. This asymmetry was interpreted by \cite{Bilous2019ApJ} as evidence for NS multipolar magnetic fields. Later on, \cite{Kalapotharakos2021ApJ} analysed the same spectroscopic and timing data for PSR J0030+0451 and interpreted it as off-centred dipole configuration with addition of quadrupolar magnetic field.

Off-centred dipole configurations are also found in other astrophysical objects. In our own solar system, the magnetic field of Mercury can be described as an off-centred dipole with a shift of $\approx 0.2$~$R_\mercury$ in the polar direction \citep{Anderson2012JGRE}.
Off-centred dipoles could result from a dynamo process operating in one hemisphere only;
this has been suggested in the context of the Martian dynamo, for example by \cite{Yan2023PSJ}.
Single hemispheric dynamos have also been found in simulations of M-dwarf stars \citep[e.g.][]{Brown2020ApJ}.
  
The main goal of this article is to study magneto-thermal evolution for off-centred dipole configurations of NSs using a modified version of the \texttt{PARODY} code \citep{parody1, parody2,Wood2015PhRvL,Gourgouliatos2016PNAS,Igoshev2021NatAs,deGrandis2020ApJ,deGrandis2021ApJ}. We plan to answer two questions: (1) how long off-centred dipole configurations survive under the influence of the Hall evolution, and (2) what are the surface thermal patterns, lightcurves and spectra produced by these configurations. We aim to determine if it is possible to estimate the dipole shift using spectral or timing observations of NSs.

Numerical models for magnetic 
and magneto-thermal evolution of NSs have been developed over the last two decades (some of the relevant references include \citealt{Hollerbach2002MNRAS,Wareing2009AA,Wareing2010JPlPh,Pons2007AA,Pons2009AA,Vigano2012CoPhC}); for a review see \cite{Pons2019LRCA}. These models were only recently extended to three dimensions \citep{Gourgouliatos2016PNAS,deGrandis2020ApJ,Igoshev2021NatAs,IgoshevGourgouliatos2021ApJ,Dehman2023PonsMNRAS}. Addition of a third spatial dimension helps to reliably model shapes and areas of hot and cold regions at the stellar surface. Thus, for the first time the results of magneto-thermal simulations are being consistently compared with spectral and timing observations of thermally emitting NSs.

We perform simulations for crust-confined field configurations. The X-ray luminosity of magnetars combined with theoretical cooling curves of magnetised neutron stars indicates that magnetic field configurations can indeed be crust-confined, see e.g. \cite{Dehman2023MNRAS}. There is some progress towards understanding the magnetic field evolution in the core, see e.g. \cite{Castillo2017MNRAS,Bransgrove2018MNRAS,Castillo2020MNRAS,Igoshev2023MNRAS}, but the material properties of the core and exact evolution equations are a matter of active debate \citep{Goldreich1992ApJ,Glampedakis2011MNRAS,Elfritz2016MNRAS,Dommes2017MNRAS,Dommes2021PhRvD,Wood2022Univ}. 

This article is structured as follows. In Section~\ref{s:methods} we describe the basic assumptions of our magneto-thermal simulations and how spectra and lightcurves are computed. In Section~\ref{s:results} we show the results of our simulations. In Section~\ref{s:discussion} we analyse different magnetised atmospheres and discuss their impact for X-ray observations. We conclude in Section~\ref{s:conclusions}.

\section{Methods}
\label{s:methods}

\subsection{Magneto-thermal evolution}

We simulate the coupled magneto-thermal evolution of crust-confined magnetic fields using the spectral code \texttt{PARODY} \citep{parody1, parody2}. A detailed derivation of the model is given in \citet{deGrandis2020ApJ}. Here we briefly summarise the equations and boundary conditions, including the model assumptions.

The NS crust is assumed to extend for 1~km from $r_c = 9$~km to $R_\mathrm{NS} = 10$~km. We assume that the crust is spherically symmetric and hence its properties depend only on the radial coordinate.
We solve equations for magneto-thermal evolution in flat space-time, neglecting corrections due to the NS compactness. In our calculations we use a simple analytical expression for the density profile within the crust given by \cite{Cumming2004ApJ}, with electron chemical potential $\mu$ (i.e.~electron Fermi energy) in the form:
\begin{equation}
\mu (r) = \mu_0 \left[1+(1-r/R_{NS}) / 0.0463 \right]^{4/3},
\end{equation}
where $\mu_0=2.9\times10^{-5}$\,erg is the value at the top of the crust.
In what follows we express quantities in dimensionless form, with $r$ measured in units of $R_{NS}$ and $\mu$ in units of $\mu_0$.
The temperature, $T$, is measured in units of $1.00\times 10^{8}$\,K. We measure magnetic fields in units of $1.00\times 10^{12}$\, G for models A, $1.00 \times 10^{13}$~G for models B and $1.00 \times 10^{14}$~G for models C.

To preserve the solenoidality of the magnetic field $\vec \nabla \cdot \vec B = 0$, it is represented as a sum of poloidal and toroidal components, which are defined in terms of poloidal potential $\beta_p$ and toroidal potential $\beta_t$:
\begin{equation}
\vec B = \nabla \times \left[\nabla \times (\beta_p \vec r) \right] + \nabla \times (\beta_t \vec r).
\end{equation}
Each potential is expanded as series of spherical harmonics $Y_{lm} (\theta, \phi)$:
\begin{equation}
\beta_p (r, \theta, \phi) = \sum_{l=1}^{N_l} \sum_{m=0}^{l} \beta_p^{lm} (r) Y_{lm} (\theta, \phi),
\end{equation} 
where $\beta_p^{lm} (r)$ is defined on discrete grid points in radius.

We solve the magnetic induction equation for electron MHD:
\begin{equation}
\frac{\partial \vec B}{\partial t} = \mathrm{Ha} \nabla \times \left[ \frac{1}{\mu^3} \vec B \times (\nabla \times \vec B) \right] - \nabla \times \left[ \frac{1}{\mu^2} \nabla \times \vec B \right] + \mathrm{Se} \nabla \left(\frac{1}{\mu}\right)\times \nabla T^2.
\label{eq:induction}
\end{equation}
Eq.~(\ref{eq:induction}) assumes
a constant relaxation time for electrons in the crust.
The unit adopted for time is the Ohmic timescale, $t_0 = 78.3$~Myr.
The dimensionless coefficients $\mathrm{Ha}$ and $\mathrm{Se}$ then take the values listed in Table~\ref{tab:const}. These coefficients are computed as:
\begin{equation}
\mathrm{Ha} = c \tau_0 \frac{e B_0}{\mu_0},
\label{eq:Ha}
\end{equation}
\begin{equation}
\mathrm{Se} = 2 \pi^3 k_\mathrm{B}^2 T_0^2 n_0 e \frac{c \tau_0}{\mu_0^2 B_0},
\label{eq:Se}
\end{equation}
where $c$ is the speed of light, $e$ is the elementary charge, $k_\mathrm{B}$ is the Boltzmann constant. $B_0$ and $T_0$ are magnetic field and temperature used to make the equations dimensionless.
The number density for electrons is made dimensionless using $n_0 = 2.6\times 10^{34}$~cm$^{-3}$. For a cold neutron star crust, electron scattering is described using a relaxation time 
\citep{Urpin1980SvA}, which
we take to be a constant
$\tau_0 = 9.9\times 10^{-19}$~s.

We also solve the heat equation in the form:
\begin{equation}
\frac{\mu^2}{\mathrm{Ro}} \frac{\partial T^2}{\partial t} = \nabla \cdot \left(\mu^2 \hat \chi \cdot \nabla T^2\right) + \frac{\mathrm{Pe}}{\mathrm{Se}} \frac{|\nabla \times \vec B|^2}{\mu^2} + \mathrm{Pe} \mu (\nabla \times \vec B)\cdot \left( \frac{T^2}{\mu^2}\right),
\label{eq:thermal}
\end{equation}
where $\hat \chi$ is the thermal conductivity tensor.
This equation assumes, for simplicity,
that the heat capacity of the crust is proportional to $\mu^2T$.
In practice, provided that the nondimensional parameter $\mathrm{Ro}$ is sufficiently large, the left-hand side of eq.~(\ref{eq:thermal}) is negligible, so the precise form of the heat capacity does not affect the evolution on the timescales of interest.
The values of dimensionless constants are computed as:
\begin{equation}
\mathrm{Pe} = \frac{3}{4\pi} \frac{B_0}{e n_0 c \tau_0}, 
\label{eq:Pe}
\end{equation}
\begin{equation}
\frac{1}{\mathrm{Ro}} = \frac{3}{4\pi^3} \frac{\mu_0^2}{k_B T_0} \frac{1}{c^2 \tau_0^2}.
\label{eq:Ro}
\end{equation}
The numerical values of $\mathrm{Pe}$ and $\mathrm{Ro}$ used in our model are also listed in Table~\ref{tab:const}. 

\begin{table}
    \caption{Values of numerical coefficients for magnetic induction and thermal diffusion equations computed for fixed $T_0 = 10^8$~K. These constants are computed using eqs.\ \ref{eq:Ha}, \ref{eq:Se}, \ref{eq:Pe} and \ref{eq:Ro}. In model A $B_0=1.00\times 10^{12}$~G, in model B $B_0=1.00\times 10^{13}$~G, and in model C $B_0=1.00\times 10^{14}$~G.}
    \label{tab:const}
    \centering
    \begin{tabular}{clrrr}
    \hline
    Constant        & Name             & Model A & Model B & Model C  \\
    \hline
    $\mathrm{Ha}$   & Hall    & 0.49    & 4.91    &  49.12 \\
    $\mathrm{Se}$   & Seebeck & 5.21    & 0.52    &  0.052 \\
    $\mathrm{Pe}$   & Peclet  & $6.44\times 10^{-7}$ & $6.44\times 10^{-6}$ & $6.44\times 10^{-5}$ \\
    $\mathrm{Ro}$   & Roberts & 3584.42 & 3584.42 & 3584.42 \\
    \hline
    \end{tabular}
\end{table}

The thermal conductivity tensor describes how heat transfer is inhibited across magnetic field lines. In index notation it has the form:
\begin{equation}
\chi_{ij} = \frac{\delta_{ij} + \mathrm{Ha} B_i B_j /\mu^2 - \mathrm{Ha} \epsilon_{ijk} B_k / \mu}{1 + \mathrm{Ha}^2 |\vec B|^2 / \mu^2},
\end{equation}
where $\delta_{ij}$ is the Kronecker symbol and $\epsilon_{ijk}$ is the Levi--Civita symbol.

Different terms in eqs.~(\ref{eq:induction}-\ref{eq:thermal}) have the following meanings. The first term in magnetic induction equation corresponds to the non-linear Hall evolution with timescale $\approx 1/\mathrm{Ha}$ shorter than the Ohmic timescale. The second term 
describes isotropic diffusion of the magnetic field, resulting from the finite resistivity of the NS crust. The last term is the Biermann battery term corresponding to possible magnetic field generation due to large thermal gradients in the crust. In our simulations, which have no imposed external heating, the influence of this term is small. The first term in the heat eq.~(\ref{eq:thermal}) describes anisotropic temperature diffusion; the second term is the heating produced by electric currents; the last term is the counterpart of the Biermann battery, and is negligible in our calculations.

In our simulations we have the following boundary conditions. We assume that magnetic field is expelled from the core (or never formed in the core) at the early stages of the evolution.
The core is therefore modelled as a perfect conductor, with $\beta_p = 0$ and $\frac{\mathrm{d}}{\mathrm{d}r}(r\beta_t) = 0$ at $r = r_c$.
We also model the region outside the star as a vacuum, with a potential magnetic field,
which implies that $\beta_t = 0$ and
\begin{equation}
\frac{\mathrm{d} \beta_p^{lm}}{\mathrm{d} r} + \frac{l+1}{r} \beta_p^{lm} = 0
\label{eq:vacuum}
\end{equation}
at $r = R_\mathrm{NS}$.
We note that Eqs.~(\ref{eq:induction})--(\ref{eq:thermal}) and our boundary condition (\ref{eq:vacuum}) neglect the curvature of spacetime produced by the star's magnetic field.
In reality, the effect of this curvature is small but not entirely negligible \citep[e.g.][]{Pons2019LRCA}.
Although we neglect this effect here,
it is fully taken account of in our calculations of X-ray spectra and lightcurves, as described in the following sections.

Our initial conditions for the magnetic field satisfy the boundary conditions and are written for a poloidal potential representing an off-centred dipole expanded using spherical harmonics. All details are summarised in Appendix~\ref{a:initial conditions}. Typically we use spherical harmonics up to degree 10 to represent large dipole shift of $d=0.5$, and only up to degree 4-6 to represent smaller shifts.
Our simulation setup is summarised in Table~\ref{t:models}. We vary the direction of the dipole shift, fraction of the shift considering $d / R_\mathrm{NS} = 0.1$, 0.3 and 0.5, strength of magnetic field and include toroidal magnetic fields in some of our simulations. We shift the dipole in the polar direction (models B1-P, B3-P and B5-P), equatorial direction (models B1-E, B3-E and B5-E) and in the diagonal direction combining shift in the equatorial and polar directions (B3-D1 and B3-D2). In this latter case we consider two models with positive and negative $d$-values, because the Hall evolution could affect these cases differently (see Appendix~\ref{A:symmtries}).
We also evolve a model with no dipolar shift for comparison (models B0-0, C0-0). We compute two models with an addition of dipole toroidal magnetic field: B0-0-tor and C0-0-tor. The toroidal magnetic field is described as $l=1$, $m=0$ harmonic only with a radial profile following \cite{Gourgouliatos2016PNAS}:
\begin{equation}
\beta_t^{10} (r) = C_t (r_\mathrm{NS} - r) (r - r_c), 
\end{equation}
where $C_t$ is a normalising constant which controls the fraction of energy in the toroidal component and $r_c$ is the crust-core boundary. In both models we assume that the energy of toroidal magnetic field coincides with the energy of poloidal magnetic field.

\begin{table}
    \caption{Description of simulation setups. Amount of dipole shift is shown as $d/R_\mathrm{NS}$ value. P stands for the shift in polar direction, E is for shift in equatorial direction and D1, D2 for a shift in combined polar and equatorial direction. $B_d$ is the dipole moment; the strength of radial magnetic field at the NS pole depends on the dipole shift and can be computed as $B_r = 2 B_\mathrm{d} / (1 - d)^2$. $\mathcal E_t = 0.50$ means that the energy of toroidal magnetic field is equal to the energy of poloidal magnetic field, i.e.\ it is half of the total magnetic energy. If $\mathcal E_t$ is not specified it means that the initial toroidal magnetic field is absent. }
    \label{t:models}
    \centering
    \begin{tabular}{lrrrrrll}
    \hline
    Name & $\log_{10} \left(B_\mathrm{d} \right)$ &  $d/R_\mathrm{NS}$ & $\mathcal E_t$ & Direction \\
     &  (G) & & &  \\
    \hline
     A3-P    & $12.00$ &  0.30 & - & P    \\
     A5-P    & $12.00$ &  0.50 & - & P    \\
     B0-0    & $13.00$ &  0.00 & - & -    \\
     B0-0-tor& $13.00$ &  0.00 & 0.50 & - \\
     B1-P    & $13.00$ &  0.10 & - & P    \\
     B3-P    & $13.00$ &  0.30 & - & P    \\
     B5-P    & $13.00$ &  0.50 & - & P    \\
     B1-E    & $13.00$ &  0.10 & - & E    \\
     B3-E    & $13.00$ &  0.30 & - & E    \\
     B5-E    & $13.00$ &  0.50 & - & E    \\
     B3-D1   & $13.00$ &  0.30 & - & D1   \\
     B3-D2   & $13.00$ &  0.30 & - & D2   \\
     C0-0    & $14.00$ &  0.00 & - & -    \\
     C0-0-tor& $14.00$ &  0.00 & 0.50 & - \\
     C3-P    & $14.00$ &  0.30 & - & P    \\
     C5-P    & $14.00$ &  0.50 & - & P    \\
     \hline
    \end{tabular}
\end{table}

For the temperature, we assume fixed core temperature, with $T (r_c) = 10^8$\,K.
Although not realistic, this helps to disentangle the thermal evolution produced by the crust itself from that arising from core cooling.
Our results here are thus a first iteration at solving this problem, but later work should ideally implement more realistic core temperatures. The surface boundary condition accounts for cooling by photon emission from the surface.
This is the dominant cooling mechanism during the later stages of neutron star evolution, when the neutrino luminosity drops below the photon luminosity. According to the simplified model by \cite{Page2004ApJS} this transition occurs at $\approx 10$~Kyr for (non-magnetised) envelopes composed of light elements, and at $\approx 100$~Kyr for envelopes composed of heavy iron-like elements.
The heat flux at the upper boundary of the model is therefore $\sigma_{SB}T_s^4$,
where $\sigma_{SB}$ is the Stefan--Boltzmann constant and $T_s$ is the surface temperature.
Following \citet{Gudmundsson1983ApJ},
we relate $T_s$ to the temperature at the upper boundary, $T_b$, via
\begin{equation}
\left(\frac{T_s}{10^6\; \mathrm{K}}\right)^2 = \left(\frac{T_b}{10^8\; \mathrm{K}} \right).
\label{e:tstb}
\end{equation}
In dimensionless form, the boundary condition at $r=R_\mathrm{NS}$ is therefore
\begin{equation}
- [\hat \chi \cdot \nabla T]_r = 10^{-9}\frac{R_\mathrm{NS}}{c\tau}\mathrm{Se}\,\mathrm{Pe}\,T.
\end{equation}
The relation~(\ref{e:tstb}) neglects the effect of magnetic fields on the heat flow through the star's atmosphere.
We consider the influence of magnetised atmospheres in Section~\ref{s:magnetised_atmospheres}.

We run our simulations using the following resolution: 96 grid points in radial direction and spherical harmonics up to degree 80. We use the surface thermal maps produced by our three-dimensional numerical simulations to produce X-ray spectra and lightcurves, as described in the next section.

\subsection{Simulations of soft X-ray spectra}

We intend to simulate soft X-ray spectra produced by NSs. While we solve magneto-thermal evolution in the flat spacetime, we model spectra and lightcurves taking into account effects of General Relativity. We describe such spectra as a sum of two black-body components following the technique developed by \cite{Yakovlev2021MNRAS}. Unfortunately, the original publication \citep{Yakovlev2021MNRAS} contained a number of typos related to numerical coefficients in effective spectral thermal flux (Yakovlev, private communication), thus we rewrite correct versions of the relevant equations here.

In order to deal with the relativistic light bending we introduce two systems of angular coordinates: two angles $\theta$ and $\phi$ describe the location of emitting surface area $dS = R_\mathrm{NS}^2 \sin \theta d\theta d\phi$ at the top of NS atmosphere (or solid surface) with respect to orientation of the original magnetic dipole; angles $\theta^\infty$ and $\phi^\infty$ are the locations at the  sphere at the observation position where the effects of general relativity are negligible. At this location we discuss instead a solid angle element $d\Omega^\infty = \sin \theta^\infty d\theta^\infty d\phi^\infty$. 

We start with the classical definition for the flux in flat space time \citep{Rybicki1985rpa}:
\begin{equation}
dF_E = I_E (\vec k, E) \cos \alpha d\Omega,
\label{eq:flat}
\end{equation}
where $I_E(\vec k, E)$ is intensity of emitted radiation in direction $\vec k$ at energy E, $d\Omega$ is the solid angle and $\alpha$ is the emission angle defined as $\cos \alpha = \vec n \cdot \vec k$, and $\vec n$ is normal to the surface. The $\cos \alpha$ term is sometimes included inside the solid angle definition.  

The total flux emitted in the direction of the observer can be computed by integrating over all solid angles which contribute to the flux (so-called visible hemisphere):
\begin{equation}
F_E = \int_\mathrm{vis} I_E (\vec k, E) \cos \alpha d\Omega.
\end{equation}
This integral is traditionally replaced by the integration over the stellar surface with $d\Omega = dS / D^2$ in flat space-time with $D$ defined as the distance between the observer and the NS. Using this replacement we introduce the definition of the \textit{effective spectral flux} $H_E$ as:
\begin{equation}
F_E = \frac{R_\mathrm{NS}^2}{D^2} H_E = \frac{R_\mathrm{NS}^2}{D^2} \int_\mathrm{vis} I_E (\vec k, E) \cos \alpha \sin \theta d\theta d\phi.  
\end{equation}
Now if we follow this analogy and rewrite eq. (\ref{eq:flat}) in the curved space-time \citep{Beloborodov2002ApJ, Poutanen2020AA} we obtain:
\begin{equation}
dF_E^\infty = I_E^\infty \cos \alpha d\Omega^\infty,
\end{equation}
where $\alpha$ is similarly defined as the angle between the normal direction and the direction of the emission at the NS surface. In order to find the total flux in the direction of the observer we have to replace $d\Omega^\infty$ with $dS$, which corresponds to appearance of the lensing factor $\mathcal D$:
\begin{equation}
\mathcal D = \frac{1}{(1 - x_g)}  \frac{d\cos \alpha}{d\cos \alpha^\infty}
\end{equation}
 in curved space-time \citep{Poutanen2020AA}. Here $x_g$ is  the compactness parameter computed as $x_g = 2 G M_\mathrm{NS} / (c^2 R_\mathrm{NS})$. A ray emitted at angle $\alpha$ to the normal direction at the NS surface bends and ends up propagating at the angle $\alpha^\infty$ far away from the NS. Thus the effective spectral flux is now written as:
\begin{equation}
H_E^\infty = \int_\mathrm{viz} I_E (\vec k, E) \mathcal{D} \cos \alpha \sin \theta d\theta d\phi.
\end{equation}
The visible hemisphere is enlarged in this case due to the light-bending. The equations allowing us to calculate the $\alpha$ and $\mathcal D$ factors are summarised by \cite{Poutanen2020AA} (their eqs. 2 and 17).

Now we assume that the local emission has a Planck spectrum with intensity
\begin{equation}
I (\vec k, E) = \frac{2}{c^2 h^3} \frac{1}{\exp(E / (k_B T_s)) - 1},
\end{equation}
where $T_s$ is the surface temperature, $k_B$ is Boltzmann's constant, and $h$ is Planck's constant.
If we replace $2 / (c^2 h^3)$ with the Stephan-Boltzmann constant
\begin{equation}
\sigma_\mathrm{SB} = \frac{2\pi^5 k_B^4}{15 c^2 h^3}    
\end{equation}
we obtain the following expression:
\begin{equation}
I_E = \frac{15 \sigma_\mathrm{SB}}{\pi^5 k_B^4} \frac{E^3 dE}{\exp(E/(k_BT_s)) - 1}.
\end{equation}

Substituting this intensity into the equation for the effective spectral flux, we obtain:
\begin{equation}
H_E^\infty = \frac{15 \sigma_{SB}}{\pi^5 k_B^4} \int_\mathrm{vis} \frac{E^3 \cos \alpha \; \mathcal{D} \sin \theta d\theta d\phi}{\exp(E/(k_B T_s^\infty)) - 1}.
\label{eq:h_e_integ}
\end{equation}
Here we also red-shift the surface temperature $T_s^\infty = T_s \sqrt{1 - x_g}$. In comparison to the expression published by \cite{Yakovlev2021MNRAS} (their eq. 4) our flux lacks a factor 16 in the denominator which appeared in \cite{Yakovlev2021MNRAS} due to a typo. Also in our case the factor $(1-x_g)^{-1}$ is included inside $\mathcal {D}$.

Eq. (\ref{eq:h_e_integ}) depends on the orientation of the NS. Thus if we see the magnetic pole (and rotational axis coincides with the magnetic pole) we can integrate over all $\phi$ and restrict $\theta$ to the interval $[0, \theta_\mathrm{max}]$ and $\alpha^\infty = \theta$. If alternatively we see the equatorial region, i.e.\ the inclination is $i = \pi / 2$, we would introduce an angle 
\begin{equation}
\cos \alpha^\infty = \sin \theta \cos \phi.    
\end{equation}
In our definition $\alpha^\infty$ corresponds to the angle $\psi$ used by \cite{Beloborodov2002ApJ} and \cite{Poutanen2020AA}. If the rotational axis does not align with the orientation of magnetic dipole, the equations becomes even more complicated; see the next section for details. 

\begin{figure*}
    \centering
    \begin{minipage}{0.49\linewidth}
    \includegraphics[width=\columnwidth]{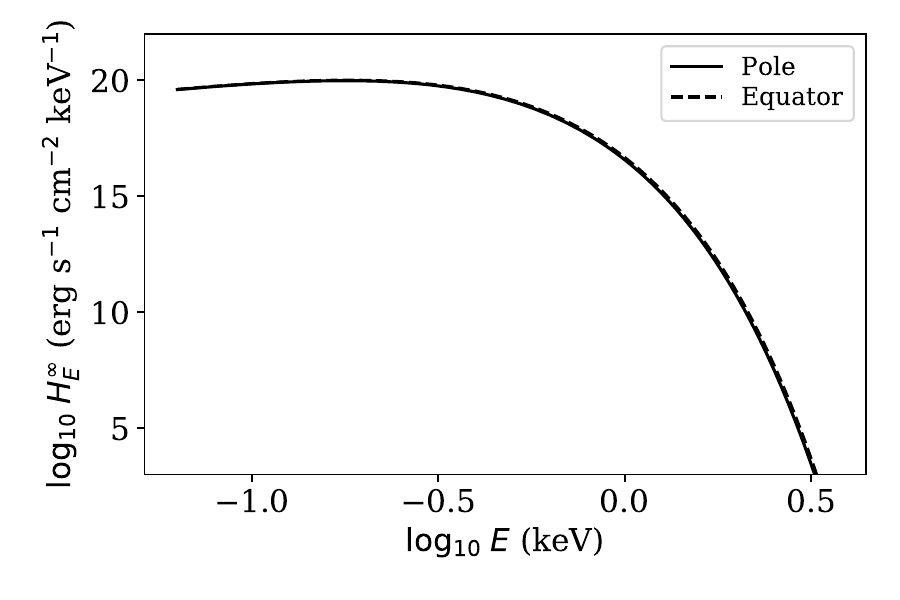}
    \end{minipage}
    \begin{minipage}{0.49\linewidth}
    \includegraphics[width=\columnwidth]{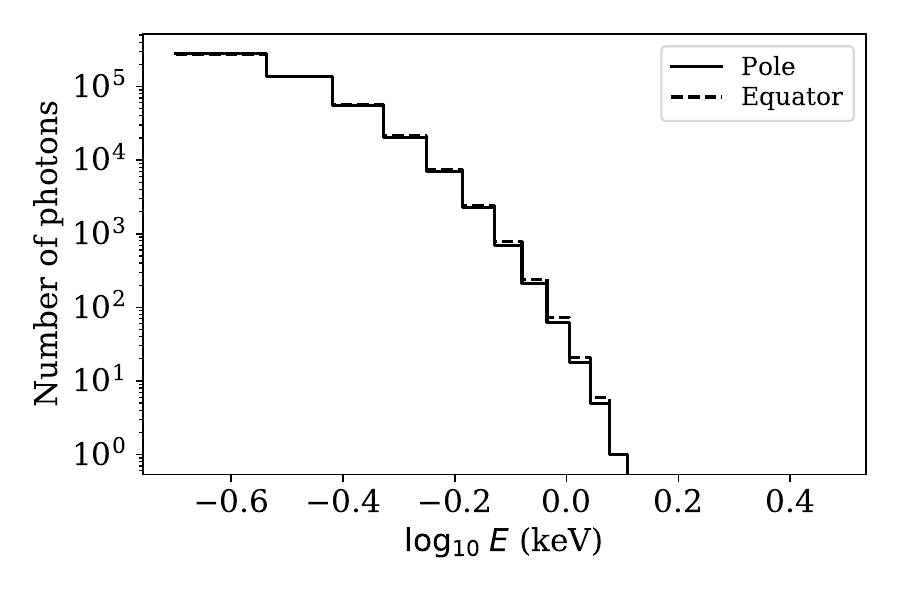}
    \end{minipage}
    \caption{Synthetic X-ray spectra generated for B5-P model at age 50~Myr assuming $M_\mathrm{NS} = 1.4$~M$_\odot$ and $R = 10$~km. Left panel shows the spectra in physical units, while right panel shows the number of photons received at different energies if we assume that the total number of received photons is fixed at  $5\times 10^5$.   }
    \label{fig:spectra_B5P}
\end{figure*}

We show the example of thermal spectra for B5-P model in Figure~\ref{fig:spectra_B5P}. This spectrum spans twenty orders of magnitude over the energy range of 0.15 to 5~keV available for modern X-ray radio telescopes such as \texttt{XMM-Newton} or \texttt{NICER}. Whole energy spectra cannot be observed and we thus concentrate only at the brightest part. In order to model spectral observations more realistically we transform the effective spectral flux into photon flux which is written as:
\begin{equation}
P_E^\infty = \frac{15 \sigma_{SB}}{\pi^5 k_B^4} \int_\mathrm{vis} \frac{E^2 \cos \alpha \; \mathcal{D} \sin \theta d\theta d\phi}{\exp(E/(k_B T_s^\infty)) - 1}.
\label{eq:p_e_integ}
\end{equation}
This flux differs from eq. (\ref{eq:h_e_integ}) by the exponent of the energy; it is proportional to $E^2$ instead of $E^3$. 
For guidance we use $10^5$~--~$5\times 10^6$ photons in our analysis, which might be achievable for the brightest NSs (\texttt{NICER} collected $10^6$ photons for a few NSs in exceptionally long exposures, see e.g. \citealt{Bogdanov2019ApJ}). We show this spectrum in the right panel of Figure~\ref{fig:spectra_B5P}. We see that only a limited part of the thermal spectrum can be probed in X-ray observation. 

If the surface temperature is uniform and does not depend on location we obtain:
\begin{equation}
H^\mathrm{BB}_E (T_\mathrm{BB}) = \frac{15 \sigma_{SB}}{\pi^4 k_B^4} \frac{1}{(1 - x_g)} \frac{E^3  \;  }{\exp(E/(k_B T_\mathrm{BB}^\infty)) - 1}.
\label{eq:BB}
\end{equation}
The factor $(1 - x_g)^{-1}$ effectively describes a fraction of area visible at the NS surface. In the case $R_\mathrm{NS} = 10$~km and $M_\mathrm{NS} = 1.4$~M$_\odot$ the value of $x_g = 0.416$ and we are able to see $(1 - x_g)^{-1} = 1.71$, i.e.\ we are able to see photons from more than a single hemisphere (which would correspond to 1). This eq. (\ref{eq:BB}) also contained a typo in the article by \cite{Yakovlev2021MNRAS}; the denominator should not have any numerical factors.

It was suggested by \cite{Yakovlev2021MNRAS} to fit the resulting spectra using a sum of two black-bodies model. We re-examine this procedure and come up with the following algorithm. We compute the photon flux eq. (\ref{eq:p_e_integ}) on the energy grid ranging from 0.2 to 3~keV assuming 85~eV energy resolution (33 separate energy bins) corresponding to \texttt{NICER} parameters and very similar to \texttt{XMM-Newton}. We fit these spectra using a two black-body model:
\begin{equation}
H_E^\mathrm{2BB} = s_1 H_E^\mathrm{BB} (p_1 T_\mathrm{eff}) +   s_2 H_E^\mathrm{BB} (p_2 T_\mathrm{eff}).   
\end{equation}
The value of $T_\mathrm{eff}$ is computed from the complete thermal map as:
\begin{equation}
T_\mathrm{eff} = \sqrt[4]{\frac{L}{4\pi \sigma_\mathrm{SB} R_\mathrm{NS}^2}}.    
\end{equation}
The total luminosity is found as:
\begin{equation}
L = \sigma_\mathrm{SB} R_\mathrm{NS}^2 \int_{4\pi} T_s^4 (\theta, \phi) \sin \theta d\theta d\phi.
\end{equation}
We restrict the total number of photons $N_\mathrm{ph}$ for synthetic spectra. The total number of photons emitted by a black-body is scaled as:
\begin{equation}
N^\mathrm{BB}_\mathrm{ph} = N_\mathrm{ph} \frac{L^\mathrm{BB}}{L}, 
\end{equation}
where $L^\mathrm{BB}$ is the luminosity of a black-body, computed as:
\begin{equation}
L^\mathrm{BB} = \sum_{i=1}^{N} 4 \pi R_\mathrm{NS}^2 \sigma_\mathrm{SB} s_i (p_i T_\mathrm{eff})^4,   
\end{equation}
where $s_i$ is the relative area of the $i$-th hot region with its relative temperature of $p_i$. This way we are able to reproduce the total luminosity of the source and get rid of the linear dependence between $s_1$ and $s_2$ in the case of sum of two black-bodies.
In order to calculate the parameters of sum of two black-bodies we use the C-statistics \citep{Cash1979ApJ} routinely employed in spectral analysis with \texttt{XSPEC} \citep{Arnaud1996ASPC}. This statistics is simply the likelihood ratio test written for the Poisson distribution of photons.
We minimise the C-statistics using the \texttt{scipy} library for two models: (1) single black-body model and (2) sum of two black-bodies. After we perform the minimisation we compare the values of statistics for both models. We select the two black-body model only if it provides a significantly better fit ($\Delta C > 2.7$).

To put things in perspective, we perform a test using the surface temperature distribution of \cite{Potekhin2003ApJ,Beznogov2021PhR} for an iron surface with $\log_{10} T_\mathrm{b} = 7.1730$, $B_\mathrm{p} = 10^{14}$~G and $N_\mathrm{ph} = 5\times 10^5$. In this case we obtain $s_1 = 0.10$, $s_2 = 0.37$ and $p_1 = 1.36, p_2 = 1.15$ and $\Delta C = 16$. These parameters are very different from ones found by \cite{Yakovlev2021MNRAS}. This difference might appear because our fit covers only the brightest part of the spectrum which could be probed in X-ray observations. It is interesting to note that if we decrease the magnetic field by three orders of magnitude, keeping the number of received photons the same, our new spectrum can easily be described with a single black-body model. We get a significant discrepancy though between $T_{\rm eff}$ and the temperature estimated in the fit as $1.18T_{\rm eff}$ while $s=0.52$. This discrepancy completely disappears if we assume that the NS has a uniform surface temperature distribution with $T=T_{\rm eff}$. This discrepancy appears because our fit for a single black-body cannot describe all details of the temperature distribution and focuses instead on the brightest part, which covers only a fraction of the NS.

We create a \texttt{Python} package \texttt{Magpies} which allows us to compute lightcurves and X-ray spectra as well as to fit synthetic X-ray spectra with single and two black-body models. The package is freely available\footnote{https://github.com/ignotur/magpies}. Detailed description of the package with example usage will follow shortly in a separate publication.

\subsection{Simulations of soft X-ray lightcurves}

In order to simulate the lightcurves we use an approach similar to calculations of spectra. Namely, we integrate the Planck function over all energies:
\begin{equation}
I(\vec k) = \frac{15 \sigma_\mathrm{SB}}{\pi^5 k_B^4} \int_0^\infty \frac{E^3 dE}{\exp{[E/(K_B T_s)] - 1}} = \frac{\sigma_\mathrm{SB} T_s^4}{\pi}.
\end{equation}
We further incorporate relativistic corrections and integrate over the visible hemisphere:
\begin{equation}
F^\infty (\chi, i, \Phi) = \frac{R_\mathrm{NS}^2}{D^2} \frac{\sigma_\mathrm{SB}}{\pi} \int_\mathrm{vis} (T_s^\infty)^4 \mathcal{D} \cos \alpha  \sin\theta d\theta d\phi.
\end{equation}
We compute this integral numerically and define a lightcurve as follows:
\begin{equation}
\frac{F (\chi, i)}{F_\mathrm{mean}} = 2\pi \frac{F^\infty}{\int F^\infty d\Phi}, 
\end{equation}
where $\Phi$ is the rotational phase. This expression depends on the star's rotational orientation defined by three angles: (1) $\chi$ is the obliquity angle between the original magnetic dipole and rotational axis, (2) $\Phi$ is the rotational phase, and (3) $i$ is the inclination of the observer with respect to the rotational axis. In order to compute $\cos \alpha^\infty = \vec o \cdot \vec R$ we need to write vectors of orientation. In this expression, $\vec o$ is vector pointing towards the observer and vector $\vec R$ is the position at the NS surface with respect to its rotational axis. We write these matrices assuming that the NS has a particular orientation in a Cartesian coordinate system. Namely, the rotational axis coincides with the $z$-axis. In the case of zero rotational phase $\Phi = 0$, vectors pointing towards the observer and magnetic pole are in $x-z$ plane. In this case the scalar product is:
$$
\cos \alpha^\infty = \vec o \cdot \vec R = (\sin i\cos \Phi, \sin i \sin \Phi, \cos i) \cdot \hspace{1cm}
$$
\begin{equation}
\hspace{2.3cm} \cdot \left[\begin{array}{ccc}
\cos \chi  & 0 & \sin \chi \\
0          & 1 & 0 \\
-\sin \chi & 0 & \cos \chi \\
\end{array}\right] \left(
\begin{array}{c}
\sin \theta \cos \phi \\
\sin \theta \sin \phi \\
\cos \theta \\      
\end{array}
\right).
\end{equation}
We derive this equation as follows. An element of NS surface with temperature $T_s$ has coordinates $\theta$ and $\phi$ with respect to the original dipole. We rotate this location by $\chi$ around the $y$-axis. Further we introduce the observer $\vec o$ which is rotated by angles $i$ and $\Phi$ around the rotational axis.
Angle $\cos \alpha^\infty$ is used to calculate $\alpha$ and $\mathcal {D}$-factors. Calculations of $\cos \alpha^\infty$, $\alpha$ and $\mathcal D$ needs to be repeated for each unique phase $\Phi$ and rotational orientations $\chi$, $i$.

\section{Results}
\label{s:results}

\subsection{The simplest models: B0-0 and C0-0}
\label{s:basic}

\begin{figure*}
    \centering
    \includegraphics[width=1.99\columnwidth]{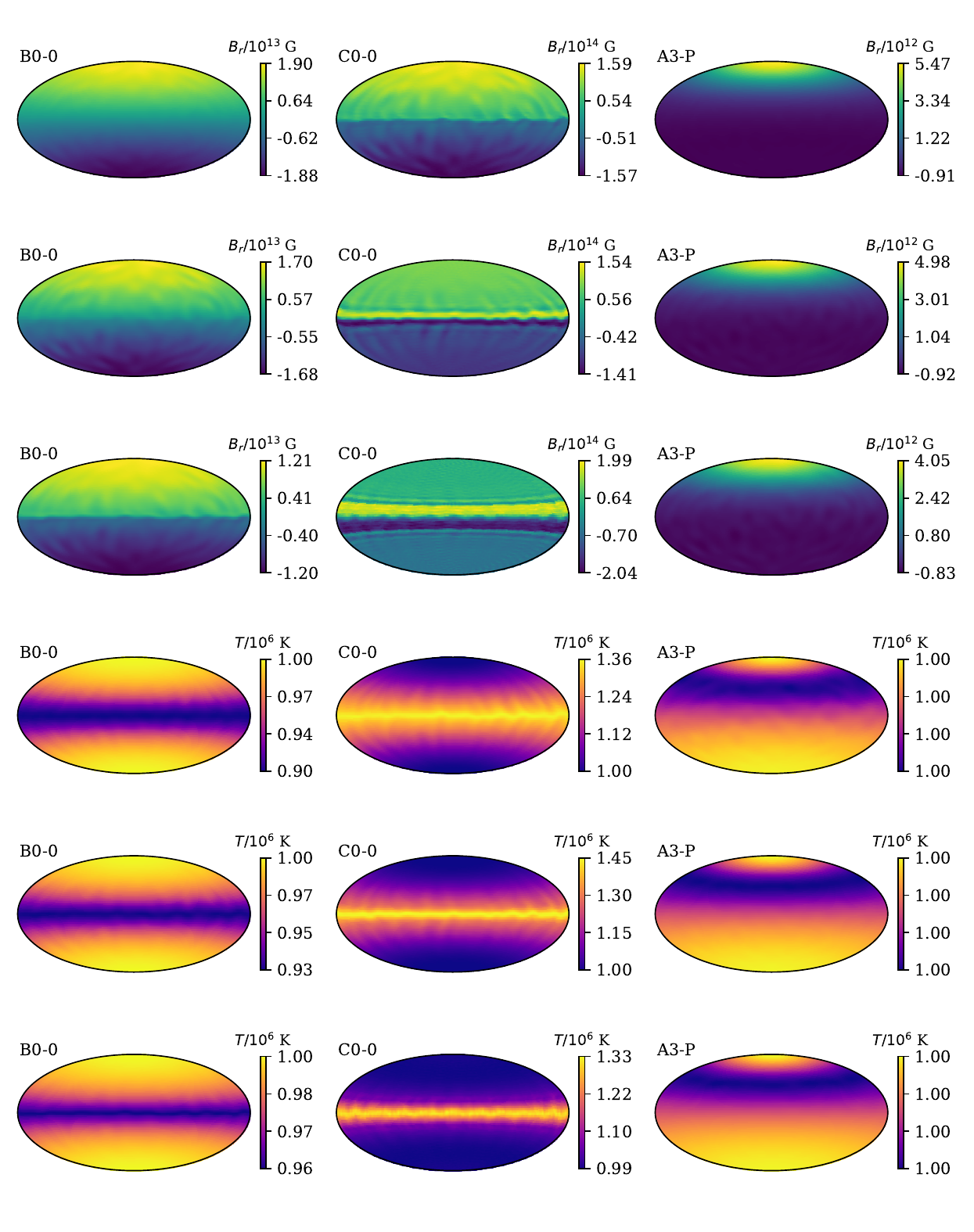}
    \caption{Results of magneto-thermal simulations for dipolar magnetic fields of different strengths (B0-0 and C0-0 models) as well as model with dipole shift of $d / R_\mathrm{NS} = 0.3$ (A3-P). The top three panels show radial magnetic fields at the top of the crust, while the lower three panels show the surface temperatures. The first row shows magnetic fields at 10~kyr, the second shows magnetic fields at 50~kyr, and the third row shows magnetic fields at 200~kyr. The same ordering also applies to the thermal maps.}
    \label{fig:basic_evol}
\end{figure*}

We show the results for B0-0 and C0-0 in first two columns of Figure~\ref{fig:basic_evol}. Similar results were presented multiple times in the literature and are shown here to simplify comparison with more advanced models. 

The magnetic field of model C0-0 is an order of magnitude stronger and it causes two effects: (1) evolution proceeds faster, and (2) energy is released in a form of heat in the equatorial region. While the external magnetic field stays dipolar-dominant in the case of B0-0 for nearly 200~kyr, this is not the case for C0-0. In this model a strong octupole component is formed and the radial magnetic field reaches its maximum values near the original equator after 200~kyr. 

The difference in thermal maps is striking. In model B0-0, the equator is cold. This is caused by decreased thermal conductivity in radial direction for equatorial region. With our choice of magnetic field $1.00\times 10^{13}$~G, the heating by internal currents plays a relatively small role and the surface thermal pattern is nearly entirely defined by thermal anisotropic magneto-diffusion process in the crust, i.e.\ by the first term on the right-hand side of eq. (\ref{eq:thermal}). Alternatively, in the case of C0-0, the equator is significantly hotter than the poles. It is caused by additional heating released by decaying electric currents concentrated in equatorial region. Looking at these thermal maps, we can foresee that X-ray variability of C-models is larger than in the case of B-models. In the future it is necessary to model this effect with more precision taking into account neutrino cooling. 

\subsection{Influence of field structure}

While studying the influence of field structure we fix the strength of magnetic field at $1.00\times 10^{13}$~G and vary the dipolar shift and its direction. This way we expect that the equator stays cold and magnetic field evolution is relatively slow. We investigate a few models with stronger magnetic fields and dipole shift in Section~\ref{s:strong_fields}. 

\subsubsection{Dipole shifted in polar direction: B1-P, B3-P and B5-P}

\begin{figure*}
    \centering
    \includegraphics[width=1.99\columnwidth]{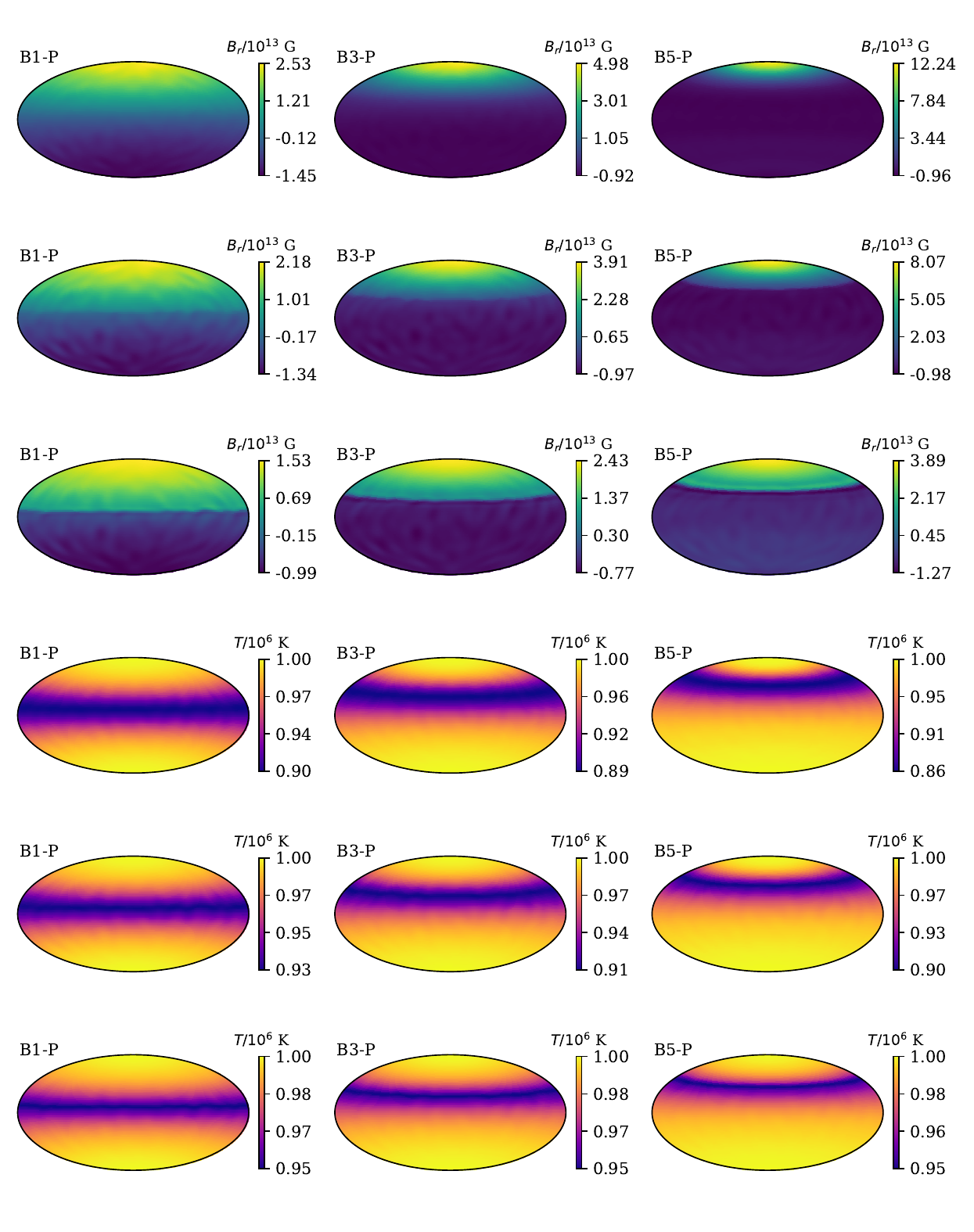}
    \caption{Results of magneto-thermal simulations for off-centred dipoles shifted in polar direction. The top three panels show radial magnetic fields at the top of the crust, while the lower three panels show the surface temperatures. The first row shows magnetic fields at 10~kyr, the second shows magnetic fields at 50~kyr, and the third row shows magnetic fields at 200~kyr. The same ordering also applies to the thermal maps.}
    \label{fig:BP_evol}
\end{figure*}

We show the results of our simulations in Figures~\ref{fig:BP_evol}. As expected, the magnetic equator shifts toward the northern hemisphere with the growth of $d$. The magnetic equator in our simulations corresponds to a colder region at the NS surface. The magnetic field configuration is relatively stable, and remains as an off-centred dipole over 200~kyr. Similarly to B0-0 model, the equatorial regions becomes sharper (faster change from positive to negative $B_r$ components) throughout the course of evolution. This is direct evidence of the Hall evolution.

In the simulations with larger $d$ value we see that the colder region is getting smaller and the surface area of two hot regions becomes unequal. Thus, the phase-resolved spectroscopy in X-ray could be an important tool to estimate the amount of shift $d$.

\subsubsection{Dipole shifted in equatorial direction: B1-E, B3-E and B5-E}

\begin{figure*}
    \centering
    \includegraphics[width=1.99\columnwidth]{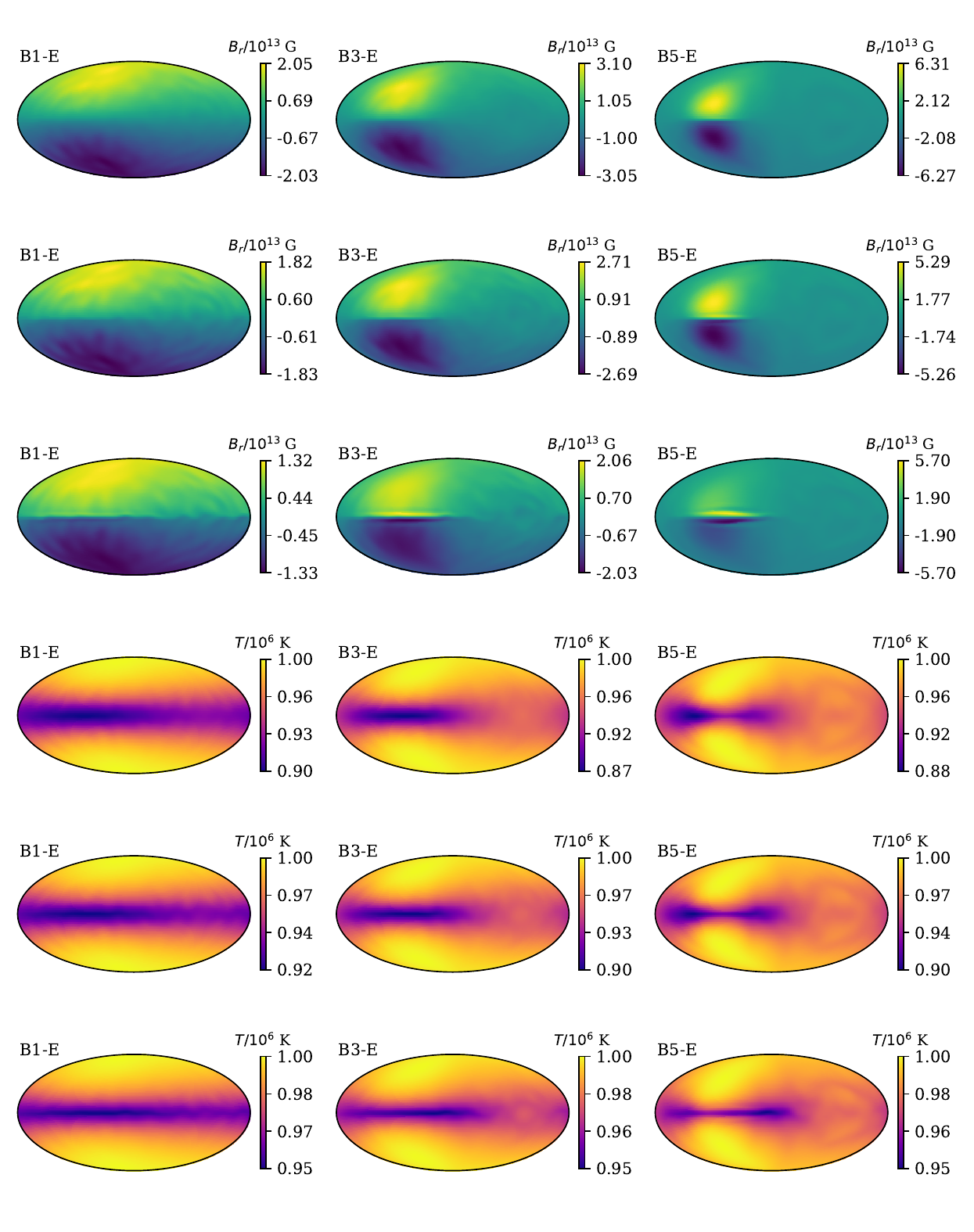}
    \caption{Results of magneto-thermal simulations for off-centred dipole shifted in equatorial direction. The top three panels show radial magnetic fields at the top of the crust, while the lower three panels show the surface temperatures. The first row shows magnetic fields at 10~kyr, the second row shows magnetic fields at 50~kyr and the third row shows magnetic fields at 200~kyr. The same ordering also applies to the thermal maps.  }
    \label{fig:BE_evol}
\end{figure*}

Figure~\ref{fig:BE_evol} shows the result of simulations with the dipole shifted in the equatorial direction. In comparison to the previous simulations, the surface magnetic field and thermal maps look quite different.

The magnetic equatorial region becomes clearly visible only at the azimuth of the shift. For larger dipole shifts $d=0.5$ the surface magnetic fields reach values twice as strong as would normally be present at the equator. As usual, the Hall effect leads to formation of sharp regions around the equator. In this particular case, this region has a small extent in azimuth and  does not cover the whole magnetic equator.

It is worth noting that in this case, the deceleration  moment from particle acceleration and electromagnetic emission will be applied non-symmetrically to the neutron star surface. This means that such a NS will probably demonstrate strong precession. Detailed numerical modelling of the magnetosphere is required in this case to understand all implications. 

As for the thermal map, for large shifts $d=0.3$ and $d=0.5$ these maps look very different from the case of a dipole shift in the polar direction. Two hot extended regions are formed above and below the colder equator which covers only a limited part of the total neutron star.

\subsubsection{Diagonal shift: B3-D1 and B3-D2}

\begin{figure*}
    \centering
    \includegraphics[width=1.99\columnwidth]{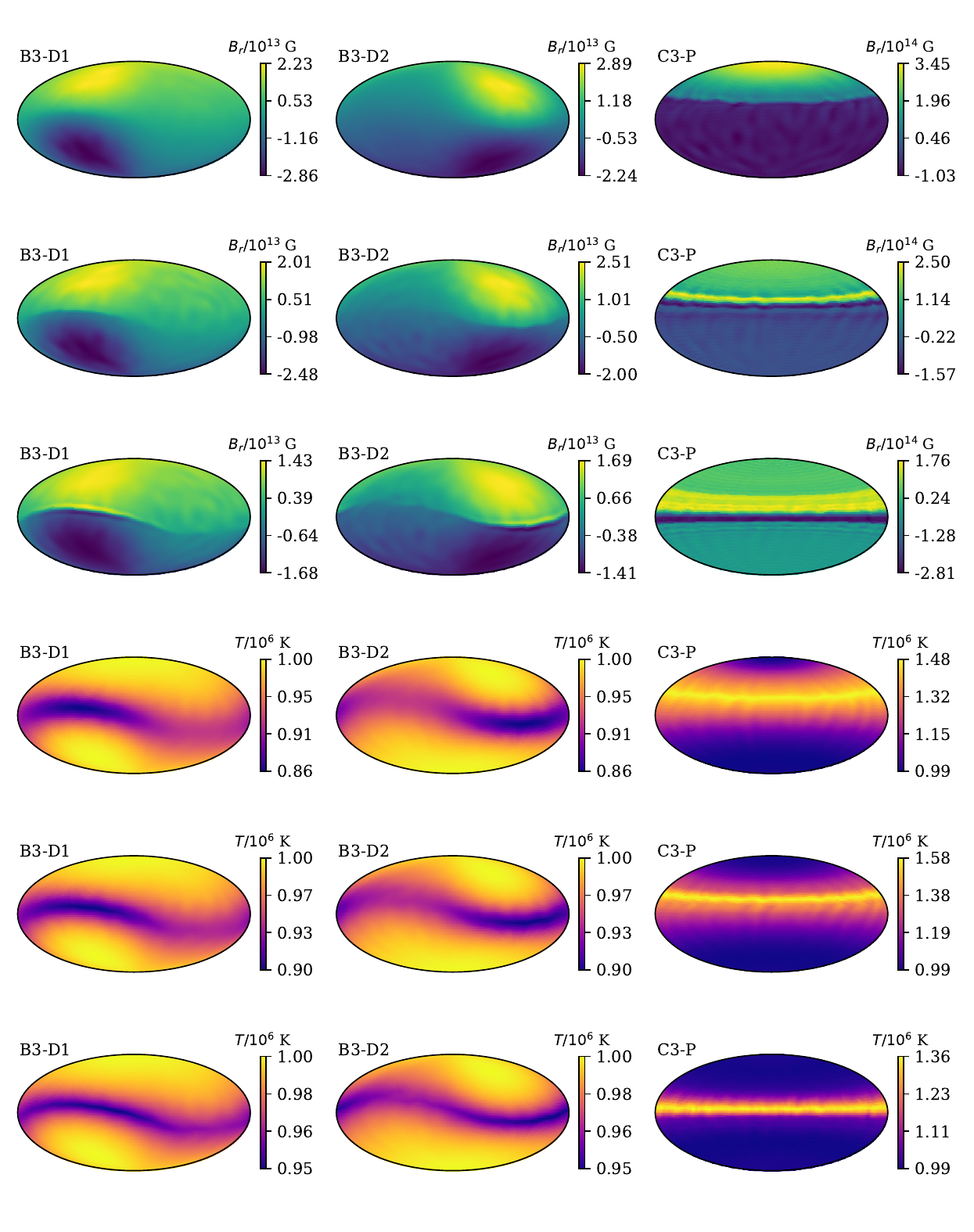}
    \caption{Results of magneto-thermal simulations for off-centred dipole shifted in diagonal directions (models B3-D1 and B3-D2) and C3-P model. The top three panels show radial magnetic fields at the top of the crust, while the lower three panels show the surface temperatures. The first row shows magnetic fields at 10~kyr, the second row shows magnetic fields at 50~kyr, and the third row shows magnetic fields at 200~kyr. The same ordering also applies to the thermal maps. }
    \label{fig:BD_evol}
\end{figure*}

While it is expected that magnetic field will evolve exactly the same way if the shift in the polar direction is positive or negative, it is not necessarily true for a shift in a diagonal direction, see Appendix~\ref{A:symmtries}. We therefore compute configurations where the dipole is shifted in positive and negative directions along the direction inclined by $45^\circ$ with respect to the orientation of the magnetic dipole axis. This results in formation of a wave-like magnetic equator at the surface of the neutron star, see Figures~\ref{fig:BD_evol}.
Overall, the magnetic field evolution is quite similar to the case of a dipole shifted in the equatorial direction, i.e.\ the equatorial region gets sharper due to the Hall evolution. 

In the thermal maps, the magnetic equatorial region corresponds to a cold spot which follows the shape of the equator. The thermal maps for simulations B3-D1 and B3-D2 look similar, and it is possible to get one from another if it is rotated by $180^\circ$ around central points in the map, i.e.\ north and south magnetic poles are inverted. 

\subsection{Weak and strong magnetic fields: A3-P and C3-P }
\label{s:strong_fields}

We compute a number of models with the same field topology, but different magnetic field strength, e.g.\ A3-P, B3-P and C3-P, see Figures~\ref{fig:basic_evol}, \ref{fig:BP_evol} and \ref{fig:BD_evol}. With respect to surface magnetic field evolution we notice the following. The speed of magnetic field evolution grows with the strength of magnetic field. While the radial magnetic field looks very similar to its original configuration after 200~kyr for A3-P it is definitely not the case for B3-P where the equator becomes noticeably sharper. In the case of C3-P, the magnetic equator shifts back to middle of NS which would corresponds to strongly decreased dipole shift i.e. to centred dipole.  

As noted in our basic comparison (B0-0 and C0-0; see Section~\ref{s:basic}), the surface thermal map is also very sensitive to the strength of magnetic field. While in the A3-P and B3-P models the equator is cold, in the C3-P model the equator becomes significantly hotter than the NS surface at the poles. This is a consequence of our thermal diffusion eq.\ (\ref{eq:thermal}) and $\mathrm{Pe} / \mathrm{Se}$ fraction. This fraction is roughly proportional to $B^2$ and describes the pace of heating by currents in the NS crust. In the case of $B_\mathrm{d} = 1.00\times 10^{14}$~G, the heating at the equatorial region is significant, explaining why the equator becomes hotter in comparison to NS poles. Comparing A3-P with B3-P we notice even though the equator is cold in both of these simulations, it is significantly colder in B3-P. This happens because our thermal conductivity tensor depends on magnetic field strength. Thus the magnetic equator is better thermally isolated from the core in B3-P than in A3-P simulation.

\subsection{Dipole shift evolution}

Strictly speaking an off-centred dipole configuration cannot be described using a finite number of spherical harmonics. However, we notice that even for large dipole shift $d=0.5$ the coefficients of the spherical harmonics expansion decay quickly with $\ell$. Specifically, $\beta_p^{\ell0} = d^{(\ell-1)}$, and $\beta_p^{40}$ is already nearly an order of magnitude smaller than the dipole. Therefore, it is possible to cut the expansion at some finite $\ell<10$. As was shown earlier \citep{Wareing2009AA,Wareing2010JPlPh} the Hall evolution leads to formation of an energy cascade where energy is distributed from dipole to small scale harmonics with $E(\ell)\propto \ell^{-2}$, see Section~\ref{s:ell_spectra} for discussion. Therefore, it quickly becomes irrelevant what (small) energy we put into small scale harmonics because more energy is delivered there via the Hall cascade.

As for the observations, the dipole moment can be measured via the pulsar spin-down. The quadrupole/octupole moments might be measured via studies of braking index, see e.g. \cite{IgoshevPopov2020MNRAS}. The braking index is defined as $n = 2 - (P \ddot P) / (\dot P^2)$ where $P$ is the spin period. The dominant component $l$ then makes the braking index to be $n=2\ell + 1$ \citep{Krolik1991ApJ}. 
It is unclear at the moment if $\ell\approx 10$ can ever be measured, perhaps via small scale features in phase-resolved X-ray spectroscopy, see e.g. \cite{Arumugasamy2018ApJ}. Thus, for practical reasons we assume that dipole shift can be described with a very limited number of spherical harmonics.   

In order to check how the shift $d / R_\mathrm{NS}$ evolves with time we introduce two measures: (1) as a fraction between quadrupolar and dipolar strengths $d_{21}$ and (2) as a fraction between octupolar and dipolar strengths $d_{31}$. We define these fractions as follows:   
\begin{equation}
d_{21} = \sqrt{\frac{5}{3}} \frac{\beta_p^{20} (R_\mathrm{NS})}{\beta_p^{10} (R_\mathrm{NS})},
\label{eq:d21}
\end{equation}
and
\begin{equation}
d_{31} = \sqrt[4]{\frac{7}{3}} \sqrt{\frac{\beta_p^{30}(R_\mathrm{NS})}{\beta_p^{10}(R_\mathrm{NS})}},
\label{eq:d31}
\end{equation}

\begin{figure*}
    \centering
    \begin{minipage}{0.49\linewidth}
    \includegraphics[width=\columnwidth]{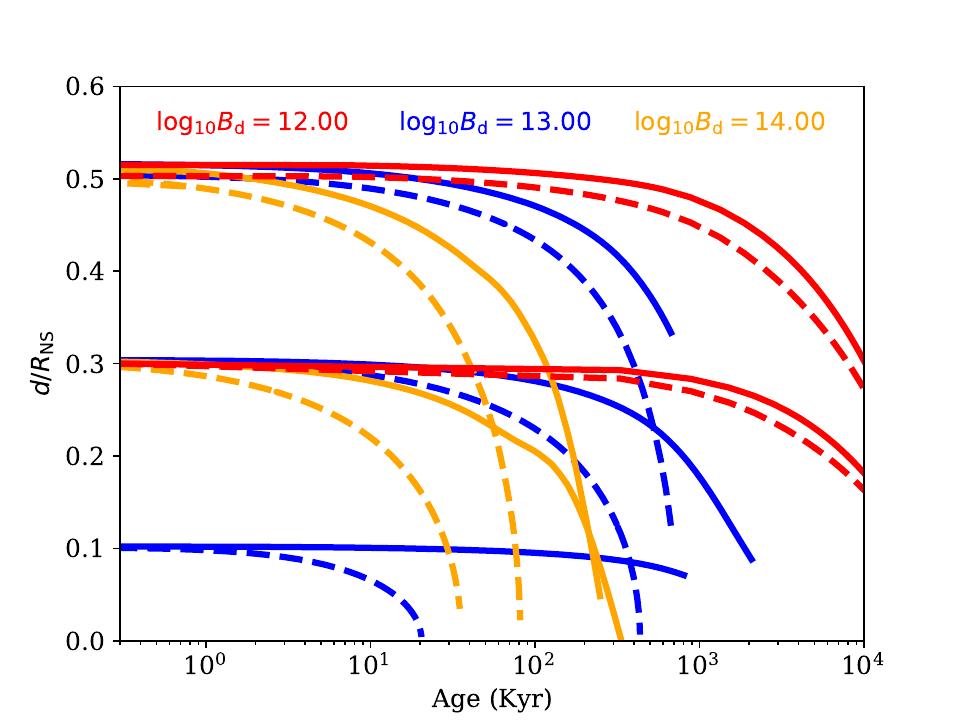}
    \end{minipage}
    \begin{minipage}{0.49\linewidth}
    \includegraphics[width=\columnwidth]{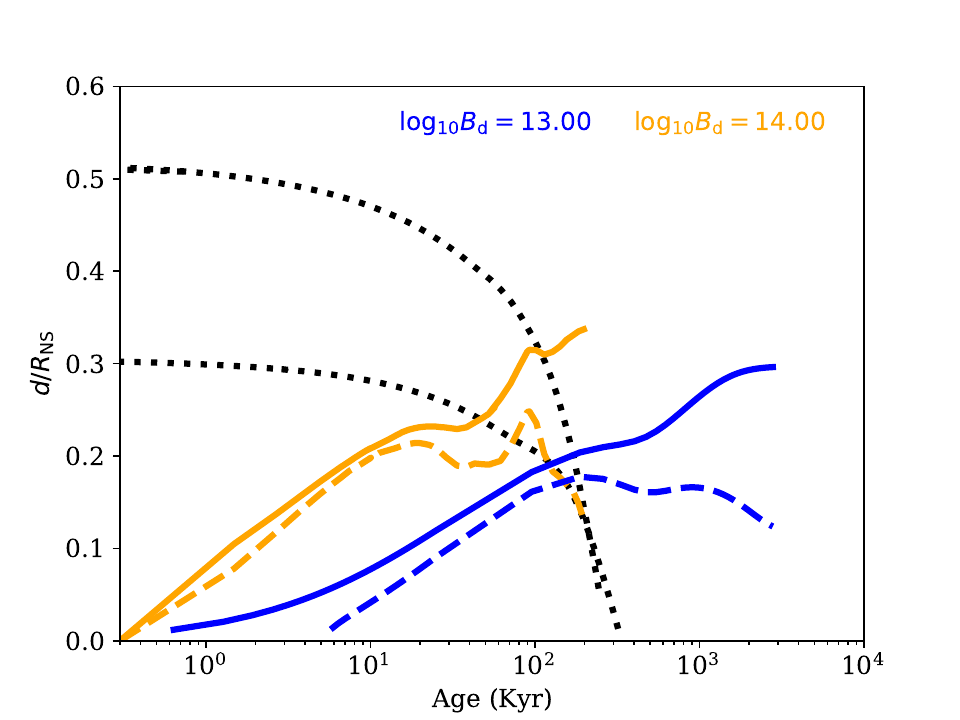}
    \end{minipage}
    \caption{Evolution of dipole shift $d$ and estimated using the dipole and quadrupole strengths ($d_{21}$ solid lines) and dipole and octupole strengths ($d_{31}$ dashed lines). Blue colour corresponds to initial $B_\mathrm{d} = 1.00\times 10^{13}$~G,  orange colour to $B_\mathrm{d} = 1.00\times 10^{14}$~G and red color to $B_\mathrm{d} = 1.00\times 10^{12}$~G. The values of dipole shift are computed at the NS surface, i.e.\ they represent properties of the external magnetic field. Left panel: configurations with no initial toroidal magnetic field. Right panel: evolution of configurations with initial toroidal magnetic field compared to C3-P and C5-P (dotted lines). }
    \label{fig:d_evol}
\end{figure*}

We show the evolution of these quantities in Figure~\ref{fig:d_evol}. We see that in all our simulations without toroidal magnetic field, the dipole shift decreases with time. This means that the magnetic field configuration evolves in the direction of a normal, centred dipole. It is interesting to notice that the octupole decays faster than the quadrupole. Thus, after a few kyr (for BP-type simulations), the estimate $d_{21} > d_{31}$. In the case of B3-P, $d_{21}$ decays from 0.3 to 0 after 400~Kyr, but $d_{31}$ becomes $0.1$ in 2~Myr. For initial condition $d=0.1$, the $d_{31}$ estimate reaches 0 within the first 20~kyr. After this $b_3$ becomes negative and thus $d_{31}$ becomes meaningless. 

It is interesting to note that the estimated dipole shift decays also when the magnetic field is stronger $B_\mathrm{d} = 1.00\times 10^{14}$~G. The dipole shift in this case decays significantly faster. For example, $d_{21}$ evolves from 0.5 to 0.3 in $\approx 200$~kyr for $B_\mathrm{d} = 1.00\times 10^{14}$~G. It is related to so-called Hall timescale of the magnetic field evolution:
\begin{equation}
\tau_\mathrm{Hall} = \frac{4\pi e n_0 L^2}{c B},
\end{equation}
where $L$ is some typical length scale associated with the electric current. Comparing models B5-P and C5-P we can expect that timescale of Hall evolution is an order of magnitude shorter in C5-P, which is visible in the estimated decay of $d_{21}$ and $d_{31}$. Similarly, the dipole shift decays significantly faster in the case of C3-P than in the case of B3-P.  

Using the results of our simulations we estimate the timescale for decay of dipole shift ($d_{21}$). We write the timescale as the following:
\begin{equation}
t_{d} = 19 \; \mathrm{Myr} \left( \frac{1.00\times 10^{12}\; \mathrm{G}}{B} \right).
\end{equation}

One implication is that older radio pulsars (ages > 10 Myr) should look more dipolar if their magnetic field is studied through polarisation in radio. It also means that even if the initial dynamo operated efficiently in a single hemisphere only (large shifts $d\approx 0.3-0.5$), these neutron stars will slowly evolve toward centred dipolar magnetic field configurations. 

It is also important to note that complicated magnetic configurations, i.e.\ dipole + quadrupole, could be interpreted in limited observations as an off-centred dipole with a shift of
\begin{equation}
d = \sqrt{\frac{5}{3}} \frac{\beta_p^{20}}{\beta_p^{10}}.
\end{equation}
We discuss the influence of toroidal magnetic field and its relation to apparent dipolar shift in Section~\ref{s:toroidal}.

It is also interesting to look back at the evolution of the B0-0 configuration, which is the initially centred dipole. It was shown by \cite{Gourgouliatos2014PhRvL} that an initial centred dipole field evolves toward generating an octupole component. Thus, we can formally compute the dipole shift using the strength of the $\beta_p^{30}$ field coefficient.
The sign of the newly forming $\beta_p^{30}$ is negative, which means that we cannot directly use eq. (\ref{eq:d31}), instead we compute the absolute value of $\beta_p^{30}$ and substitute it into this equation. 
At 10~kyr, $d_{31}=0.11$, and the octopole component is negligible. At 50~kyr $d_{21}=0.20$, and by 200~kyr it reaches values $d_{21}=0.28$~R$_\mathrm{NS}$. This effective shift does not mean that the dipole is shifted in the opposite hemisphere, but it signifies that the total magnetic field becomes more squashed in comparison to a centred dipole. Also, the $\beta_p^{20}$ field coefficient stays much smaller in comparison to $\beta_p^{30}$.

\subsection{Strong toroidal magnetic field: B0-0-tor and C0-0-tor}
\label{s:toroidal}

We run simulations for two model with large scale, crust-confined toroidal magnetic field.
A similar configuration was discussed by \cite{Igoshev2021NatAs}. 
We expand the radial surface magnetic field using the spherical harmonics and estimate the dipolar shift using eq.\ (\ref{eq:d21}) and (\ref{eq:d31}). We show the result of the evolution in the right panel of Figure~\ref{fig:d_evol}. In contrast to simulations with the poloidal magnetic field, the dipole shift grows significantly with time. It reaches the maximum value $\approx 0.3$ when the toroidal energy is equal to poloidal energy.

\begin{figure*}
    \centering
    \includegraphics[width=1.99\columnwidth]{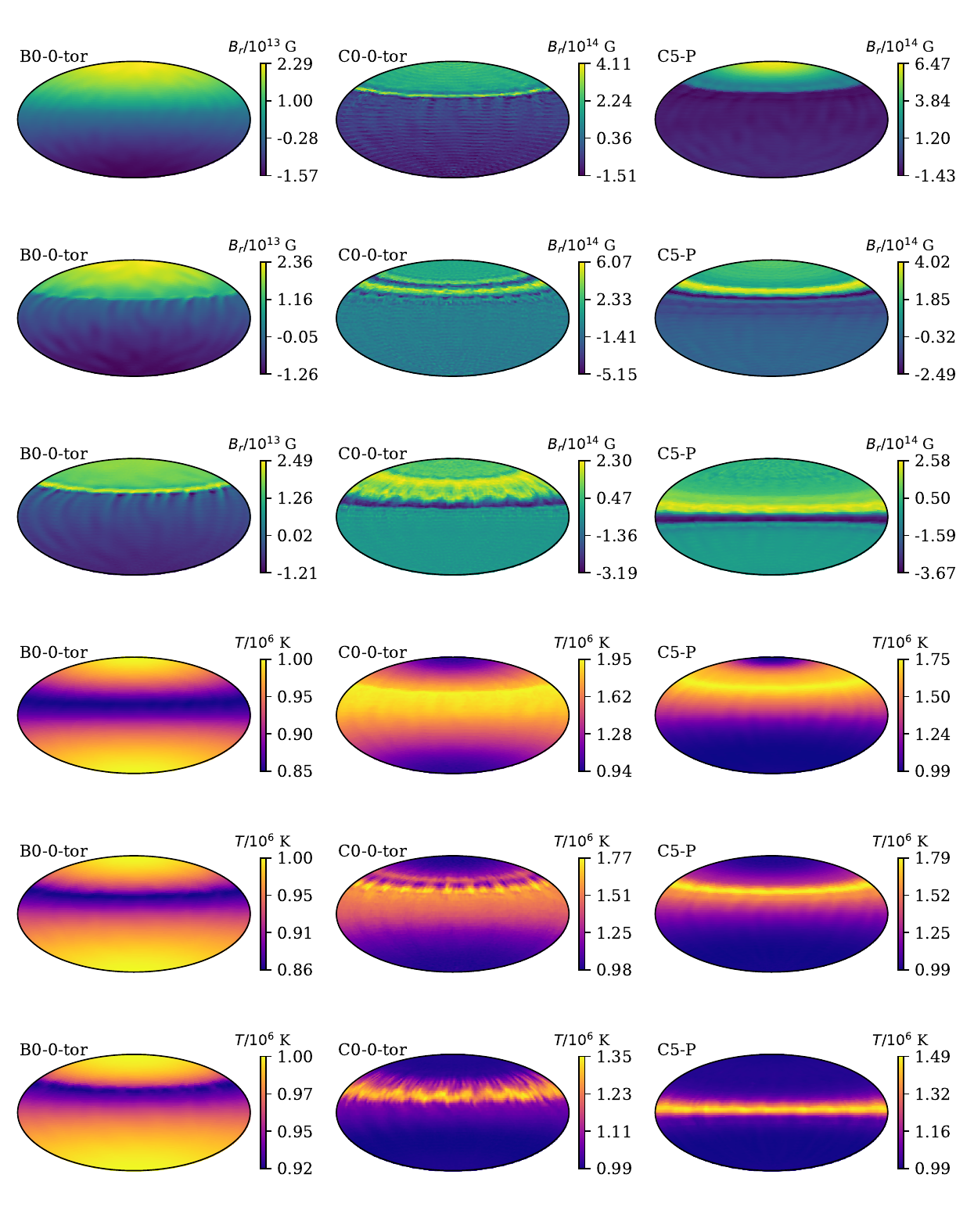}
    \caption{Results of magneto-thermal simulations for configurations with toroidal magnetic field (B0-0-tor and C0-0-tor). The last column shows the results of simulations for C5-P model. The top three panels show radial magnetic fields at the top of the crust, while the lower three panels show the surface temperatures. The first row shows magnetic fields at 10~kyr, the second row shows magnetic fields at 50~kyr, and the third row shows magnetic fields at 200~kyr. The same ordering also applies to the thermal maps.  }
    \label{fig:CP_evol}
\end{figure*}

The growth of dipole shift is easy to see in magnetic maps, see Figure~\ref{fig:CP_evol}. Even though the initial configuration B0-0-tor looks like a centred dipole, the equator becomes shifted in the northern hemisphere within the first 50~kyr. In the case of C0-0-tor, the shift already occurs within the first 10~kyr. Another interesting aspect can be noticed when we compare C0-0-tor with C3-P simulation, see Figure~\ref{fig:BD_evol}. In the latter case, the initial dipole shift of $d/R_\mathrm{NS}=0.3$ evolves towards a centred dipole during 200~kyr, which does not happen in C0-0-tor case which stays significantly off-centred even after 200~kyr.  

Thus, a configuration with strong toroidal magnetic field could be similar to the off-centred dipole configuration. It is important to highlight here a difference between C0-0-tor and C5-P. Although these configurations looks quite similar at 50~kyr, there is a noticeable difference in their thermal maps. In the case of C0-0-tor there is additional azimuthal structure in the hot region. Essentially, the hot spot breaks into multiple separate hot spots extended in the north-south direction. This evolution occurs only in this case and is caused by Hall instability \citep{GourgouliatosPons2019PhRvR} due to tearing instability of toroidal magnetic field near the surface.

\subsection{X-ray lightcurves}

\begin{figure*}
    \centering
    \begin{minipage}{0.49\linewidth}
    \includegraphics[width=\columnwidth]{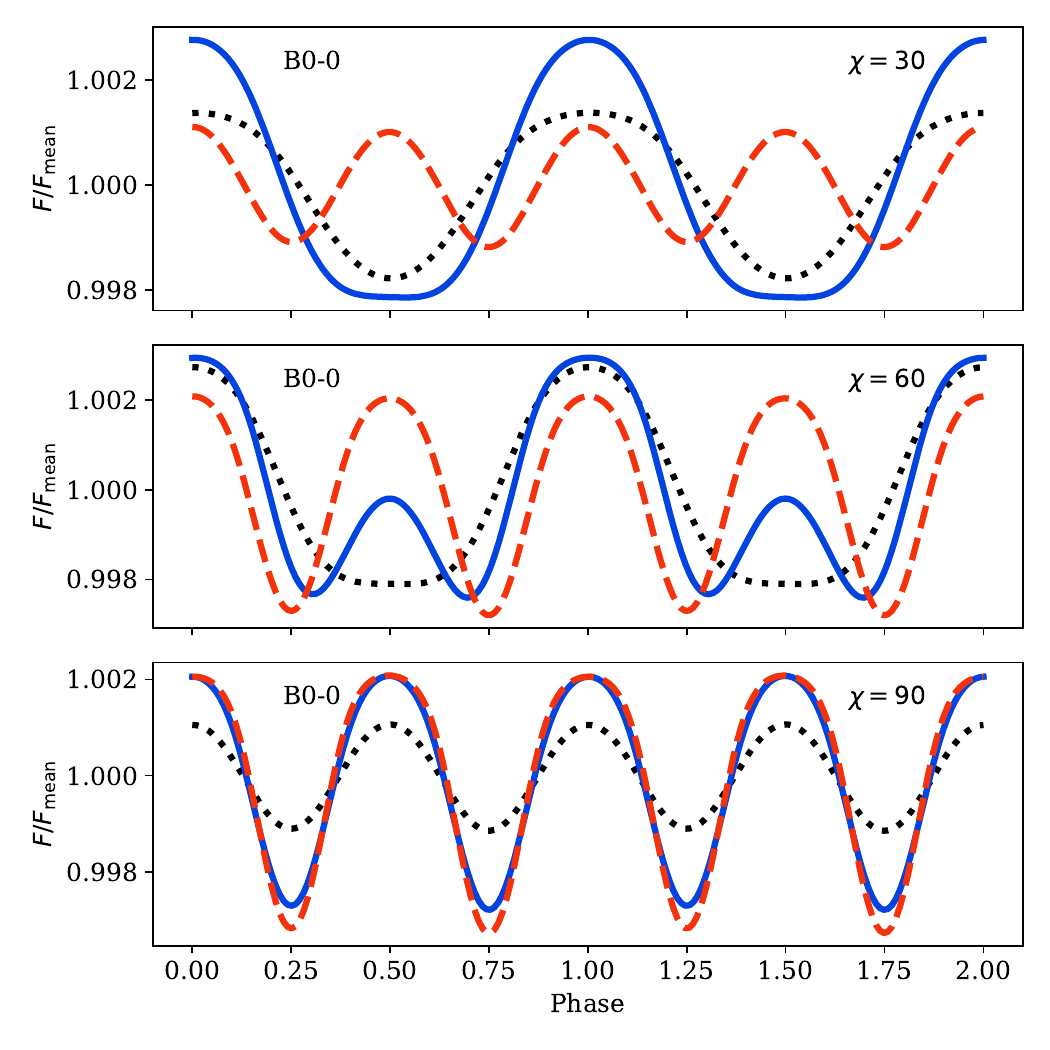}
    \end{minipage}
    \begin{minipage}{0.49\linewidth}
    \includegraphics[width=\columnwidth]{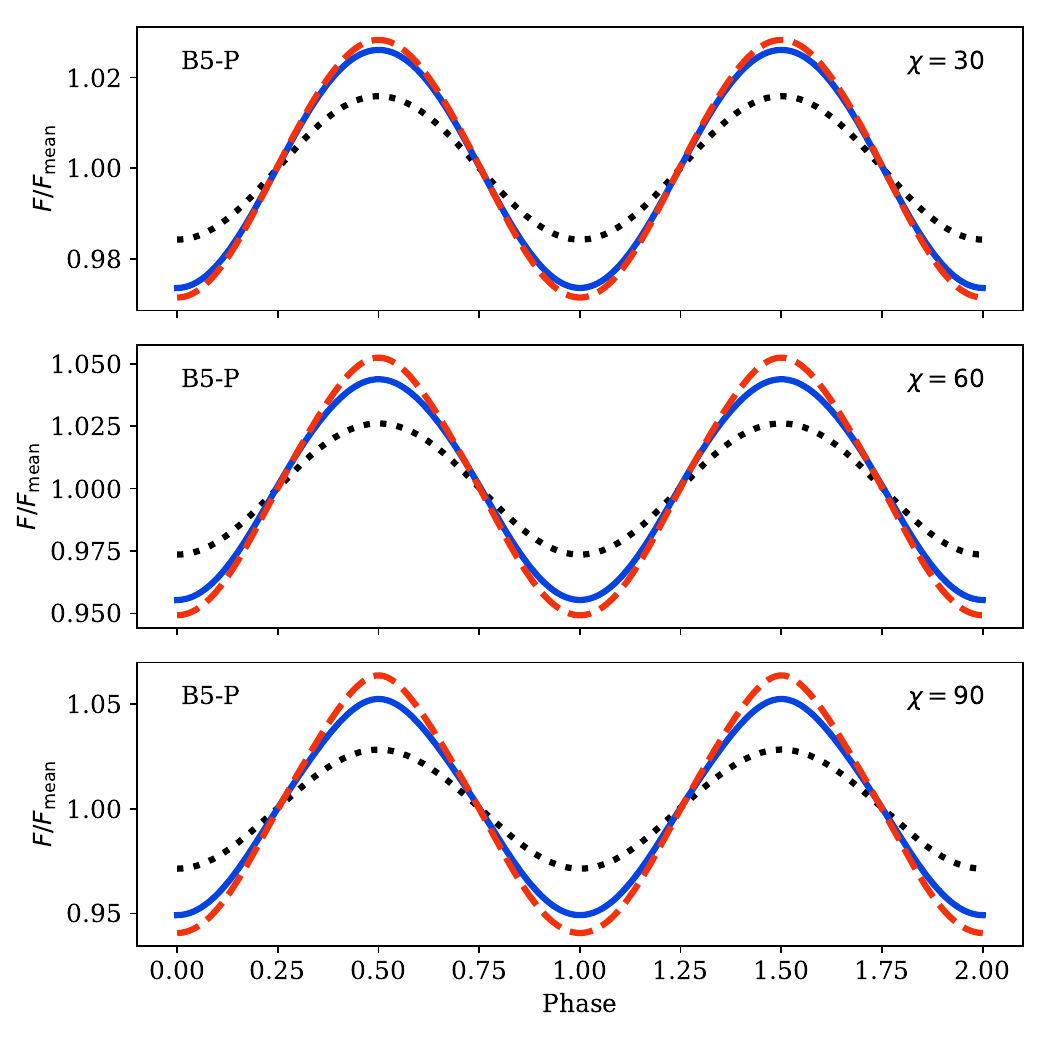}
    \end{minipage}
    \begin{minipage}{0.49\linewidth}
    \includegraphics[width=\columnwidth]{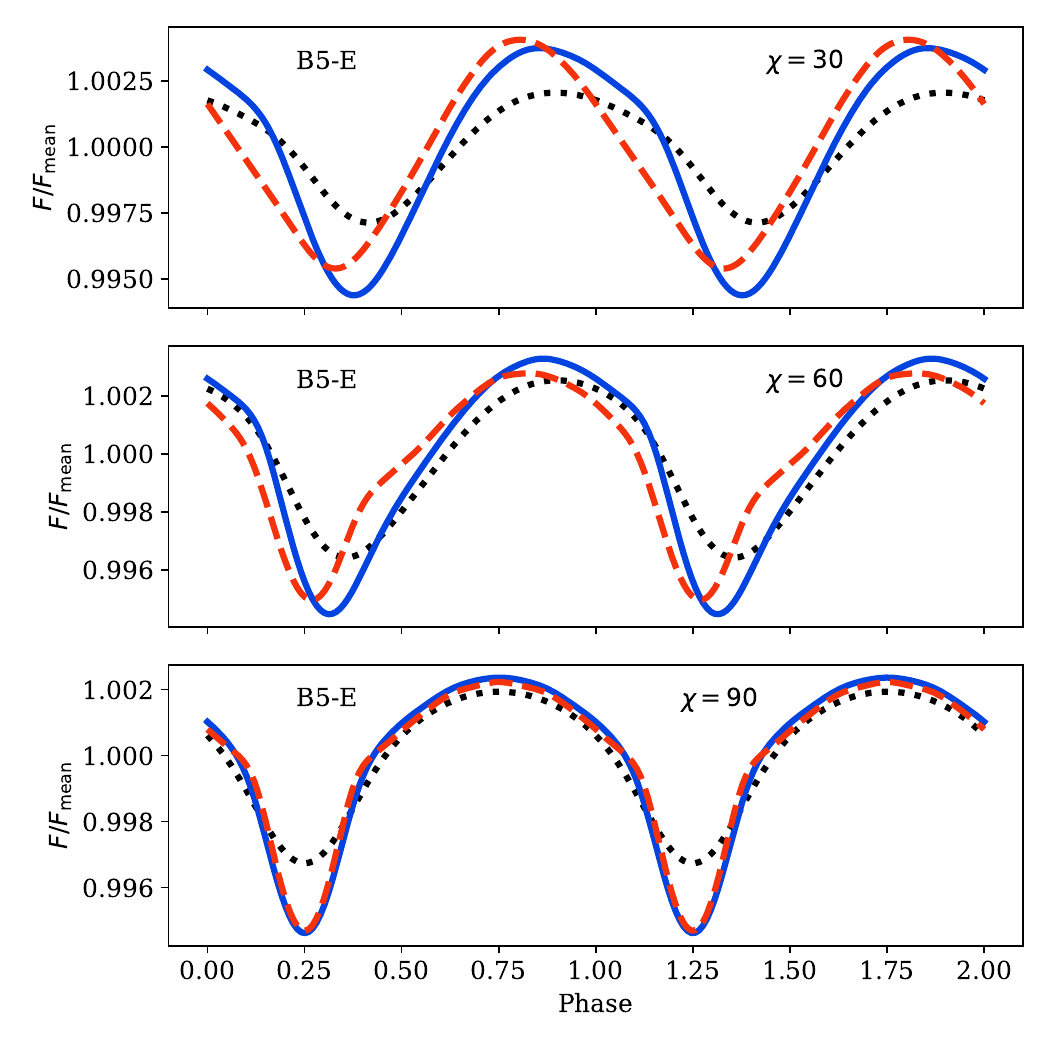}
    \end{minipage}
    \begin{minipage}{0.49\linewidth}
    \includegraphics[width=\columnwidth]{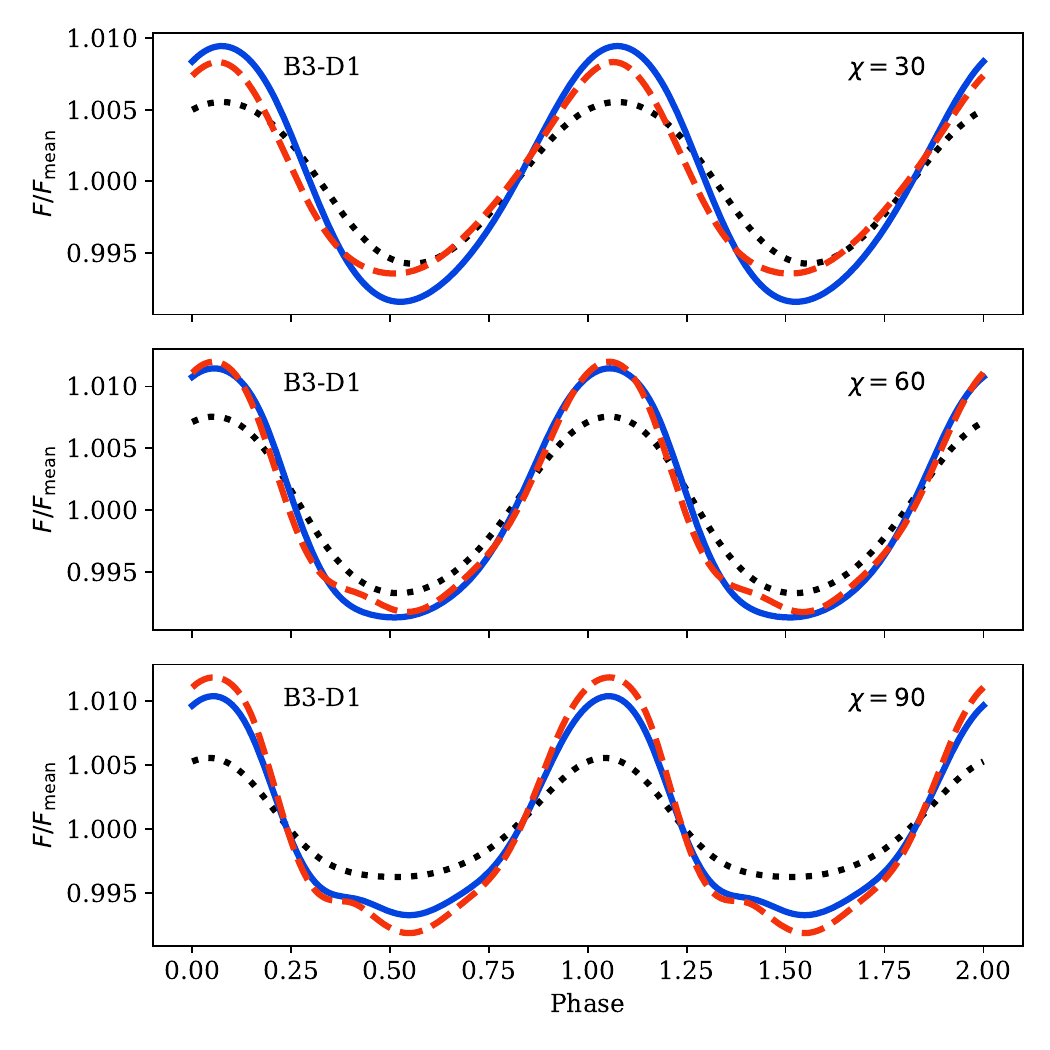}
    \end{minipage}
    \caption{Soft X-ray lightcurves for different configurations of surface magnetic field computed at 50~kyr for models B0-0, B5-P, B5-E and B5-D1. Each panel corresponds to a different obliquity angle $\chi$ between the orientation of the original magnetic dipole and the rotation axis. Lines of different colours and types correspond to different inclinations angles: black dotted line is for to $i=30^\circ$; solid blue line is for $i=60^\circ$ and dashed red line is for $i=90^\circ$.}
    \label{fig:lightcurves}
\end{figure*}

Figure~\ref{fig:lightcurves} shows the lightcurves for our models B0-0, B5-P, B5-E and B3-D1 computed at 50~kyr. We summarise maximum pulsed fraction at age of 50~kyr for all models in Table~\ref{tab:pf_and_ps}.  The lightcurves are presented for exploratory purposes only. Detailed study of a source will require lightcurves for any arbitrary orientation and age. We propose that in the future machine learning tools will be developed to classify and characterise new NS observations. These tools will be trained on the most sophisticated theoretical models. 

As we have already seen from the thermal maps and spectral properties of A and B models, NS with shifted dipole configurations typically have large hot spots which cover tens of percent of the total surface area. It means that we do not expect large amplitudes of X-ray intensity variation during a single rotation. The computed lightcurves confirm our expectations. Only in B5-P case does the amplitude (also known as pulsed fraction) reach 6~percent, while in the remaining cases it is about a few percent. Surprisingly, the B5-P model produces a simple sine-shaped lightcurve.

\begin{table*}
    \centering
    \caption{Parameters of 2BB fit for all models computed at 50~kyr. We assume $R = 10$~km and $M = 1.4$~M$_\odot$. While fitting the spectra we assume that we received $5\times 10^5$ photons. Maximum pulsed fractions (PF) are computed at 50~kyr assuming no magnetic beaming.  }
    \label{tab:pf_and_ps}
    \begin{tabular}{lrcccccccccc}
    \hline
    Model & Max PF & \multicolumn{5}{c}{Pole} &  \multicolumn{5}{c}{Equator}\\
          & (\%)   & $s_1$ & $s_2$ & $p_1$ & $p_2$ & $p_1/p_2$ & $s_1$ & $s_2$ & $p_1$ & $p_2$ & $p_1/p_2$ \\
    \hline
     A3-P    &  0.1  &  0.99  &  0.0  &  1.0  &  0.0  & - &  0.99  &  0.0  &  1.0  &  0.0  & - \\ 
     A5-P    &  0.4  & 1.0  &  0.0  &  1.0  &  0.0  & - &  0.99  &  0.0  &  1.0  &  0.0  & - \\
     B0-0    &  0.3  &  0.98  &  0.0  &  1.0  &  0.0  & - &  0.99  &  0.0  &  1.0  &  0.0  & - \\
     B0-0-tor&  5.4  &  0.8  &  0.21  &  1.02  &  0.91  &   1.11  & 0.17  &  0.79  &  1.09  &  0.99  &   1.1  \\
     B1-P    &  1.4  & 1.0  &  0.0  &  1.0  &  0.0  & - & 0.99  &  0.0  &  1.0  &  0.0  & - \\
     B3-P    &  4.3  & 1.02  &  0.0  &  0.99  &  0.0  & - &  0.98  &  0.0  &  1.0  &  0.0  & - \\
     B5-P    &  6.1  & 1.04  &  0.0  &  0.99  &  0.0  & - & 0.98  &  0.0  &  1.0  &  0.0  & - \\
     B1-E    &  0.3  & 0.98  &  0.0  &  1.0  &  0.0  & - & 0.99  &  0.0  &  1.0  &  0.0  & - \\
     B3-E    &  0.5  & 0.98  &  0.0  &  1.0  &  0.0  & - & 0.98  &  0.0  &  1.0  &  0.0  & - \\
     B5-E    & 0.5   & 0.99  &  0.0  &  1.0  &  0.0  & - &  0.99  &  0.0  &  1.0  &  0.0  & - \\
     B3-D1   &  1.0  & 0.96  &  0.0  &  1.01  &  0.0  & - &  0.98  &  0.0  &  1.01  &  0.0  & - \\
     B3-D2   &  1.0  &  0.99  &  0.0  &  1.0  &  0.0  & - &  0.98  &  0.0  &  1.0  &  0.0  & - \\
     C0-0    &  1.1  & 0.43  &  0.39  &  1.14  &  0.91  &   1.26  &  0.44  &  0.38  &  1.14  &  0.91  &   1.25  \\
     C0-0-tor&  23.6 & 0.46  &  0.21  &  1.17  &  0.91  &   1.28  & 0.48  &  0.26  &  1.15  &  0.87  &   1.33  \\ 
     C3-P    &  19.2 & 0.39  &  0.28  &  1.19  &  0.93  &   1.29  & 0.39  &  0.34  &  1.19  &  0.9  &   1.32  \\ 
     C5-P    &  34.5 &  0.31  &  0.22  &  1.28  &  0.94  &   1.37  &  0.32  &  0.29  &  1.27  &  0.88  &   1.44  \\ 
    \hline
    \end{tabular}
\end{table*}

\begin{figure*}
    \centering
    \begin{minipage}{0.49\linewidth}
    \includegraphics[width=\columnwidth]{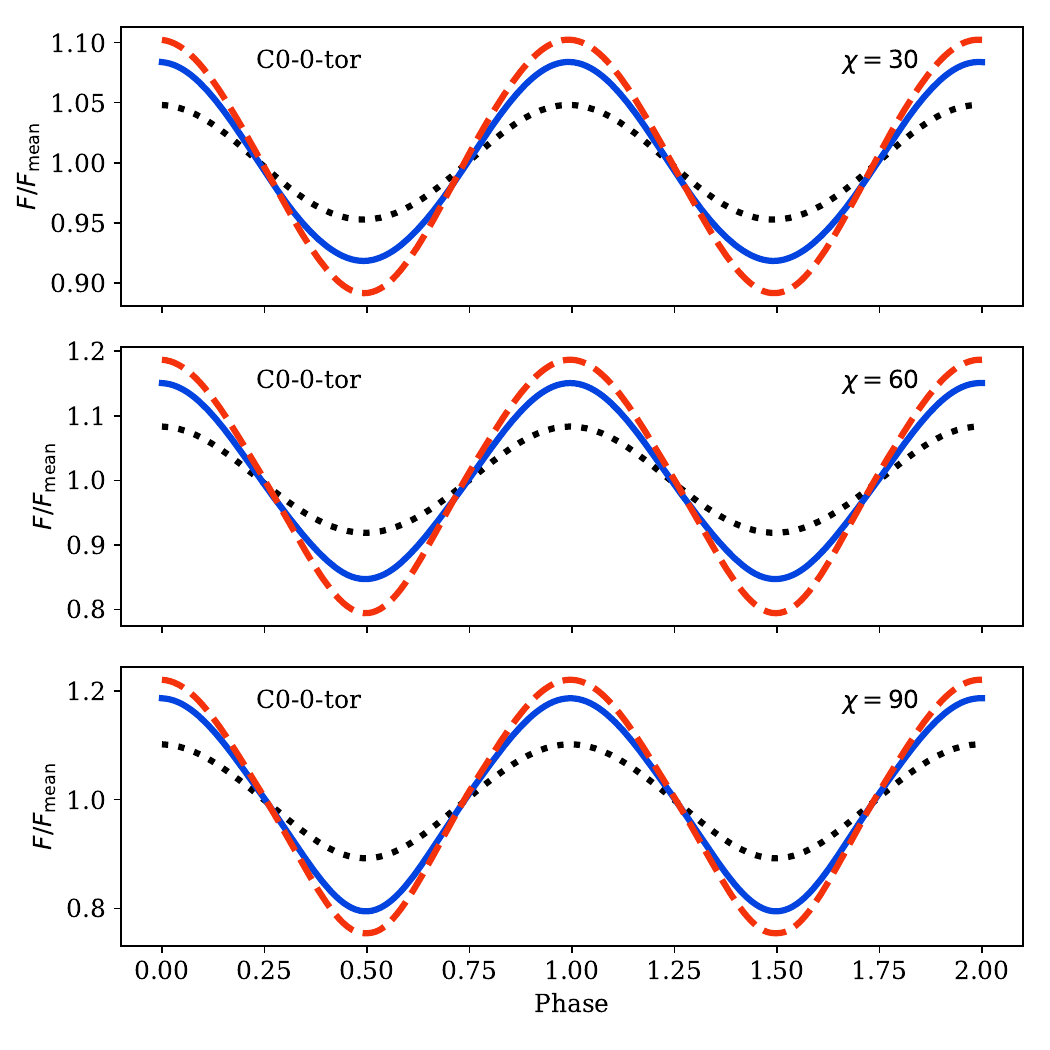}
    \end{minipage}
    \begin{minipage}{0.49\linewidth}
    \includegraphics[width=\columnwidth]{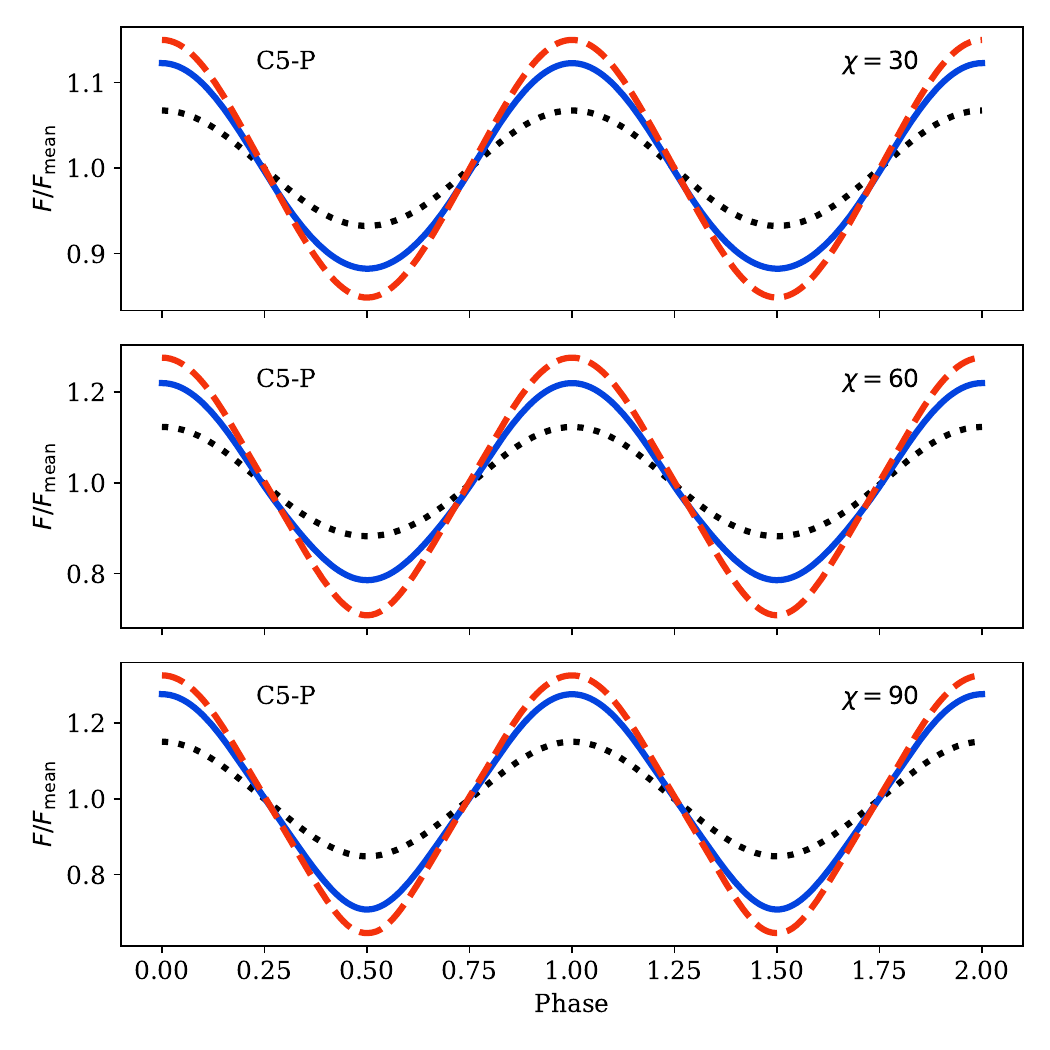}
    \end{minipage}
    \caption{Soft X-ray lightcurves for different configurations of surface magnetic field computed at 50~kyr for models C0-0-tor and C5-P. Each panel corresponds to a different obliquity angle $\chi$ between the orientation of the original magnetic dipole and the rotation axis. Lines of different colours and types correspond to different inclinations angles: black dotted line is for to $i=30^\circ$; solid blue line is for $i=60^\circ$ and dashed red line is for $i=90^\circ$.}
    \label{fig:lightcurves_Cmodels}
\end{figure*}

The most complicated (non-symmetric around the maxima) and easily recognisable shapes appear when the dipole moment has a shift in the equatorial direction (models B5-E and B3-D1). The pulsed fraction is low, and thus it would require long X-ray observations to reliably characterise such a lightcurve. In our modelling we assume no additional magnetic beaming. When we tried using beaming proportional to $\cos^2 \alpha$ we obtained more symmetrical lightcurves.

Figure~\ref{fig:lightcurves} shows the lightcurves for diagonal shift. As expected, the lightcurves are not symmetric around the maxima, but the pulsed fraction is small and only reaches $\approx 2$~percent in a favourable orientation. This happens because the hot regions are large and cover a significant part of the NS surface.

\begin{figure}
    \centering
    \includegraphics[width=\columnwidth]{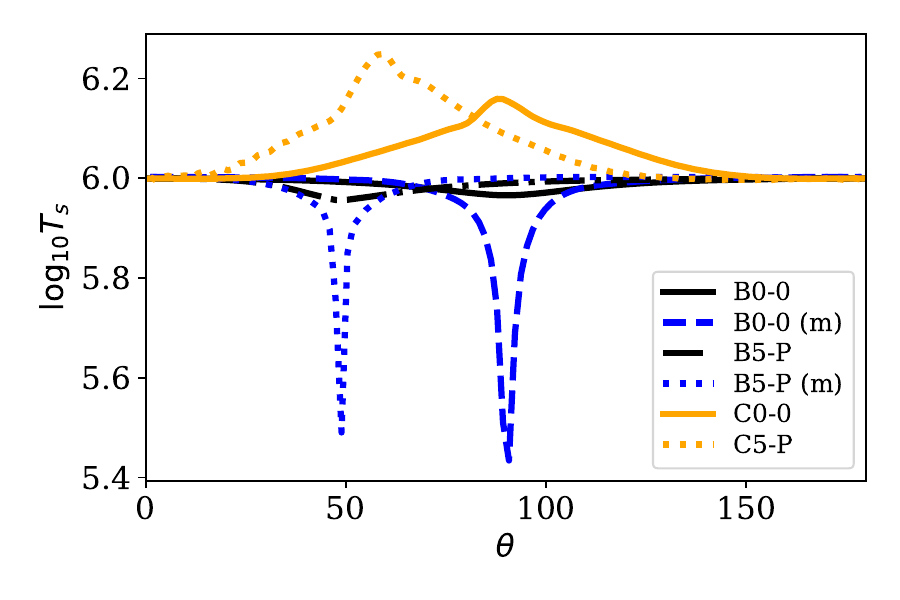}
    \caption{Temperature profile for B0-0 (black solid line), B5-P models (black dashed and dotted line), C0-0 (orange solid line) and C3-P (orange dotted line) together with profiles obtained using \protect\cite{Potekhin2015SSRv} atmospheric models with $T_b = 7\times 10^7$~K (except C3-P model). Dashed blue line is obtained using magnetic field configuration of B0-0 model and dotted blue line is obtained using magnetic field configuration of B5-P models. }
    \label{fig:atmos}
\end{figure}

We show lightcurves for C0-0-tor and C5-P models in Figure~\ref{fig:lightcurves_Cmodels}. It is easy to see that the amplitude of these lightcurves is significantly larger than in the previous cases. It happens because hot equator shifted in northern hemisphere becomes a significant source of emission and variability. We show the surface temperature profile for multiple models in Figure~\ref{fig:atmos}.  

\subsection{X-ray spectra}

We produce synthetic spectra and analyse them for models with age 50~kyr. We show parameters of two-blackbody fits in Table~\ref{tab:pf_and_ps}.
We found that for all B-models and $d$ parameters the spectra can be successfully fitted with a single black-body with $T \approx T_\mathrm{eff}$ if only $5\times 10^5$ photons are received. This happens because the temperature of colder regions in our simulations is comparable to temperature of hotter regions, see Figure~\ref{fig:BP_evol}. When more magnetised atmospheres are considered, the temperature of colder regions could be a few times smaller, see Section~\ref{s:magnetised_atmospheres} for discussion.

\begin{figure}
    \centering
    \begin{minipage}{0.49\linewidth}
    \includegraphics[width=\columnwidth]{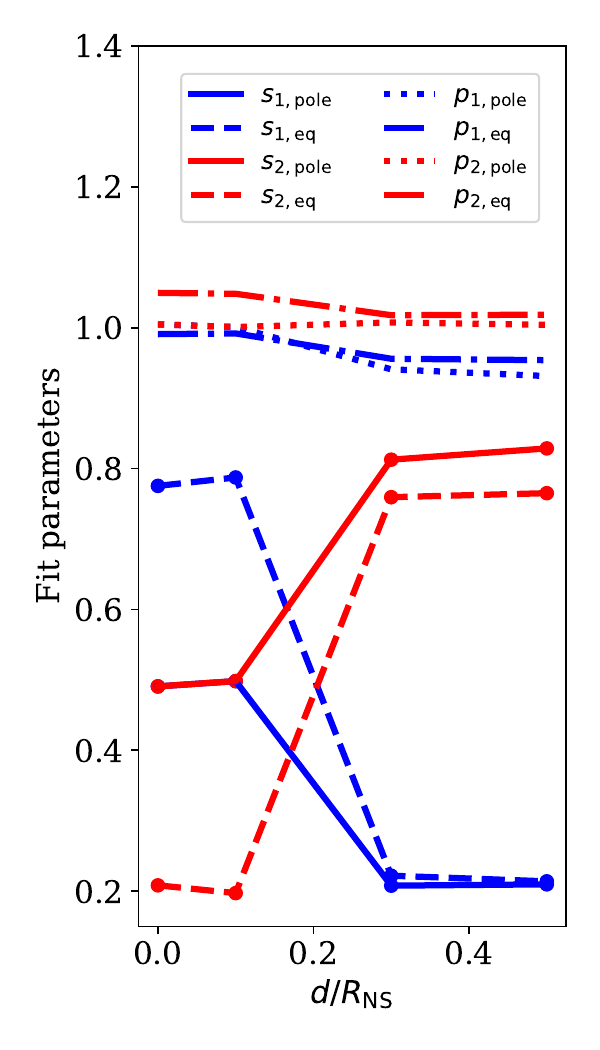}
    \end{minipage}
    \begin{minipage}{0.49\linewidth}
    \includegraphics[width=\columnwidth]{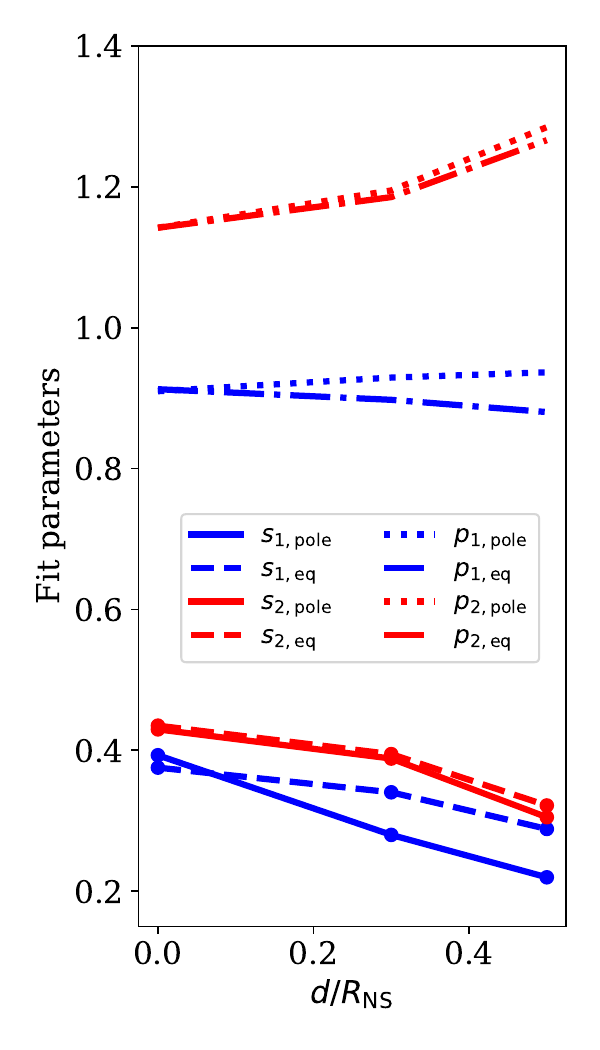}
    \end{minipage}
    \caption{Parameters of two-blackbody model for soft X-ray spectra as a function of dipole shift ($d$-value) at 50~Kyr for BP models (left panel) and CP models (right panel) assuming that we receive $5\times 10^6$ thermal photons (left panel) and $5\times 10^5$ photons (right panel).  }
    \label{fig:hot_areas_evol}
\end{figure}

We perform additional simulations increasing the number of received photons and checking if the fit for a sum of two black-bodies reaches $\Delta C = 2.7$. We found that for $5 \times 10^6$ photons we can resolve two separate black-bodies for some orientations, see left panel of Figure~\ref{fig:hot_areas_evol}. While the temperatures of two black-bodies are very similar, the surface areas differ by a factor of four and show some sensitivity to the dipole offset.

\begin{figure}
    \centering
    \includegraphics[width=\columnwidth]{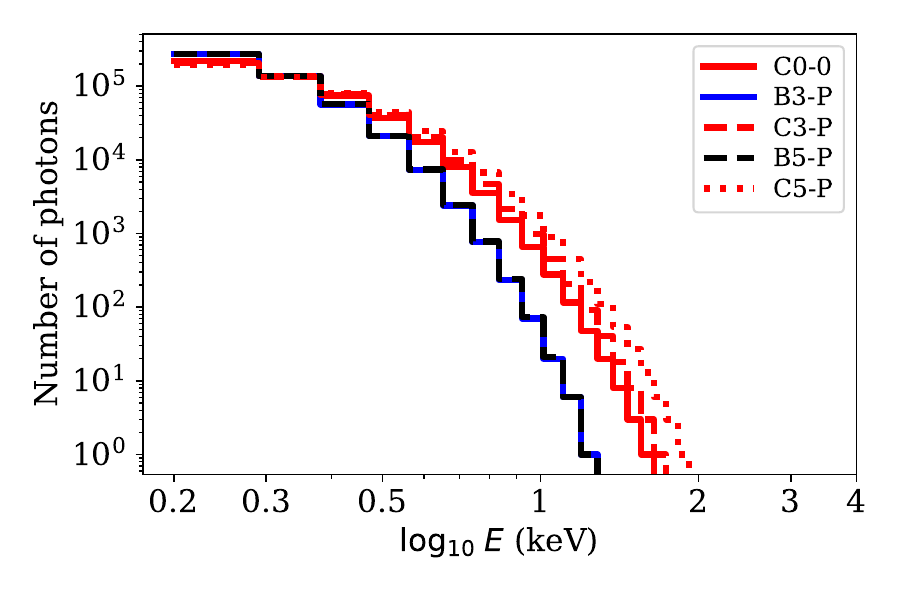}
    \caption{Synthetic X-ray spectra generated for C0-0, C3-P and C5-P models (red lines) in comparison with B3-P, B5-P models (black and blue overlapping lines) at age 50~kyr, assuming $M_\mathrm{NS} = 1.4$~M$_\odot$ and $R = 10$~km. We assume that the total number of received photons is $5\times 10^5$. }
    \label{fig:spectra_B3P_C3P}
\end{figure}

As for C-models, the situation is very different. Because the magnetic equator is hot it becomes clearly visible in X-ray spectra. At the moment we assume that we receive a similar number of photons from B3-P and C3-P NSs, which practically means that these NSs are located at different distances\footnote{NS distances are frequently relatively poorly known with uncertainty reaching a factor of a few. It happens because it is hard to measure parallaxes in X-ray due to limited angular resolution of modern X-ray telescopes.  }. We produce synthetic spectra for $i=\pi/2$ (equator on) orientation, see Figure~\ref{fig:spectra_B3P_C3P}. While these spectra approximately coincide around $200$~eV, the hot equator contributes significantly above $\approx 1$~keV. As for 2BB fits, it is practically impossible to distinguish separate hot components in B3-P model with just $5\times 10^5$~photons, but two separate components are perfectly noticeable in the case of C3-P model with $s_1\approx s_2\approx 0.36$ while $p_1 = 1.2$ and $p_2 = 0.9$. It translates into $p_1 / p_2 \approx 1.33$. The effective temperature of C3-P is $\approx 1.2$ times higher that that of B3-P.

Making a similar comparison between B5-P and C5-P we notice that the C5-P model also forms a hot equator which is located in the northern hemisphere and slowly shifts toward the central position. The temperature of the hot equator reaches $\approx 1.8$~MK\footnote{We do not model the details of NS cooling so these numbers are provided here only for exploratory purposes.}. We produce X-ray spectra for both B5-P and C5-P and show them in Figure~\ref{fig:spectra_B3P_C3P}. We can immediately notice that it is hard to distinguish between B3-P and B5-P, which coincides with results which we obtain fitting 2BB models to B3-P and B5-P. Comparing B5-P and C5-P with each other we find that these models are clearly different, with many more photons with $E\approx 1.5$~kev. It is interesting to note that there is a noticeable difference between C3-P and C5-P, with C5-P demonstrating higher temperatures.   

When we fit the synthetic X-ray spectra of the C5-P model we found the following parameters: $s_1 = 0.31$, $s_2 = 0.22$, $p_1 = 1.28$ and $p_2 = 0.94$ which means $p_1 / p_2 \approx 1.37$. It is interesting to note that in both cases $s_1 + s_2 < 1$, which is explained by the fact that a 2BB model does not fit the temperature distribution perfectly and the fit is possible only if some parameters are not realistic.  We show the dependence between parameters in Figure~\ref{fig:hot_areas_evol}.

\section{Discussion}
\label{s:discussion}

\subsection{Numerical resolution}
\label{s:ell_spectra}

\begin{figure}
    \centering
    \includegraphics[width=\columnwidth]{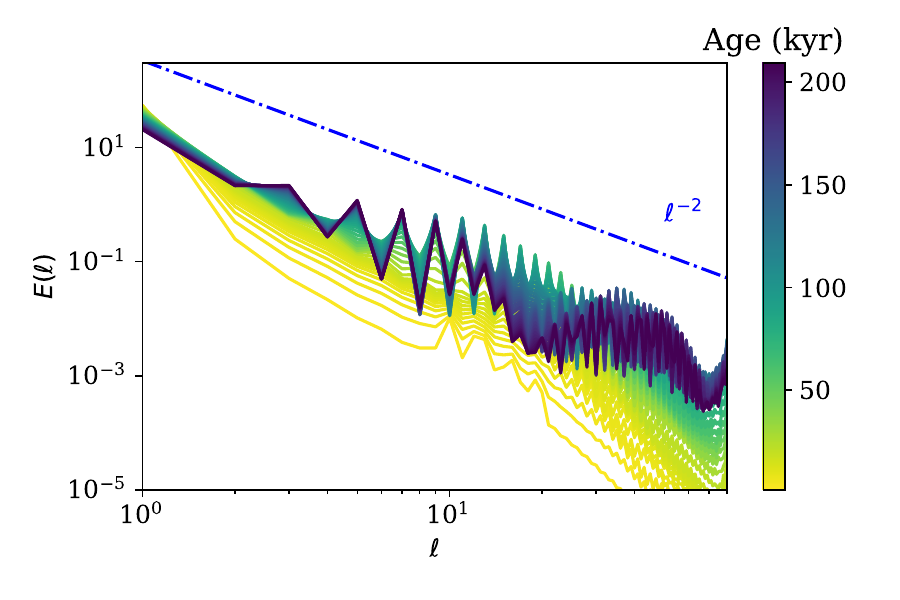}
    \caption{Evolution of magnetic energy spectra for C0-0 model. The energy spectra are computed across the NS crust and show the total magnetic energy per spherical harmonic. Each coloured line corresponds to a separate age.   }
    \label{fig:energy_spec_l}
\end{figure}

We limit the angular resolution to spherical harmonic degree and order up to 80, in order to keep the computational time manageable; the required CPU time is proportional to $\approx L_\mathrm{max} M_\mathrm{max}$. In order to check if our numerical resolution is adequate for this problem we look at the energy spectra evolution. Models with stronger magnetic fields tend to form smaller patterns faster. Therefore, we examine C0-0 model in Figure~\ref{fig:energy_spec_l}. As seen from this figure, $E(\ell)\propto \ell^{-2}$ at more advanced times. This happens due to the Hall cascade which distributes the energy from the low-$\ell$ to high-$\ell$ \citep{Goldreich1992ApJ,Wareing2009AA,Wareing2010JPlPh}.      

It is possible to see the effect of limited angular resolution because $E(\ell)$ starts growing after $\ell > 70$ at later times. The influence of this effect is very small because the energy at higher harmonics is significantly smaller ($\approx 10^2$ times less) than the energy in $\ell = 1, 2$ and $3$. The peaks at odd-$\ell$ are related to the properties of the Hall cascade since it most efficiently transfers energy to harmonics of the same parity, thus dipolar energy is most efficiently transferred to $\ell = 3, 5, 7, ...$. Some numerical noise was also injected at harmonics $10<\ell < 20$ to make the simulation less axisymmetric. Effects of limited angular resolution are significantly less noticeable in A and B simulations. 

\subsection{Influence of magnetised atmosphere}
\label{s:magnetised_atmospheres}

The role of magnetised atmospheres in cooling of magnetars was recently highlighted by \cite{Dehman2023MNRAS}. We therefore wish to explore how important magnetised atmospheres are for X-ray spectra of some of our numerical models. Our approach is very simple: first we extract the components of the surface magnetic field $B_r, B_\theta$ and $B_\phi$ from our simulations, then we compute the magnetic latitude for each point at the NS surface as:
\begin{equation}
\cos \theta' = \frac{B_r}{\sqrt{B_r^2 + B_\theta^2 + B_\phi^2}}.
\end{equation}
We then create a magnetised atmosphere following the numerical fit by \cite{Potekhin2015SSRv} and using $B_\mathrm{p} = 1.9\times 10^{13}$~G (corresponds to polar magnetic strength), $T_b = 7\times 10^7$~K, $g_{14}=1.6$ (corresponds to $R_\mathrm{NS} = 10$~km and $M_\mathrm{NS} = 1.4$~M$_\odot$) and magnetic latitude $\theta'$. 

It is important to note here that magnetised atmospheres are normally computed for the fixed value of $T_b$ at densities $\approx 10^{10}$~g~cm$^{-3}$. In strongly magnetised NS the deep crust temperatures even at larger densities $\approx 10^{13}$~g~cm$^{-3}$ are not uniform. Thus, it is unclear if the current magnetised atmospheres model can reliably describe the surface temperature distribution. That is why we do not perform the complete modelling and restrict only to simplified modelling which follows the assumptions about fixed $T_b$. This way we get complimentary results to our non-magnetised atmospheres described by eq. (\ref{e:tstb}). 

In comparison to our previous results the temperature is significantly cooler at the magnetic equator, see Figure~\ref{fig:atmos}. It is unclear at the moment how physical this feature is (see e.g.\ discussion in the paper by \citealt{Yakovlev2021MNRAS}).

\begin{figure}
    \centering
    \begin{minipage}{0.49\linewidth}
    \includegraphics[width=\columnwidth]{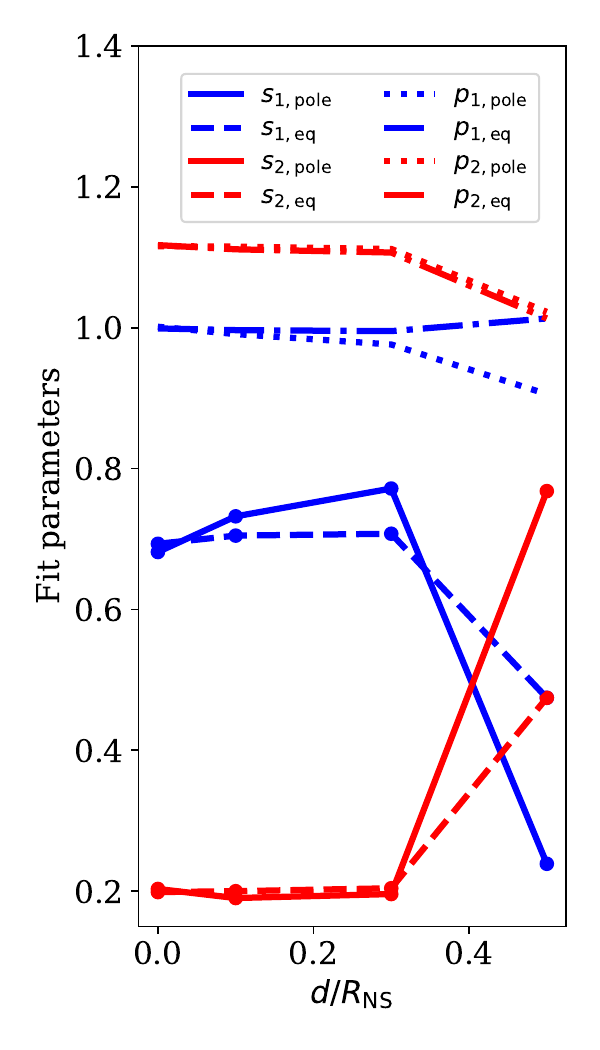}
    \end{minipage}
    \begin{minipage}{0.49\linewidth}
    \includegraphics[width=\columnwidth]{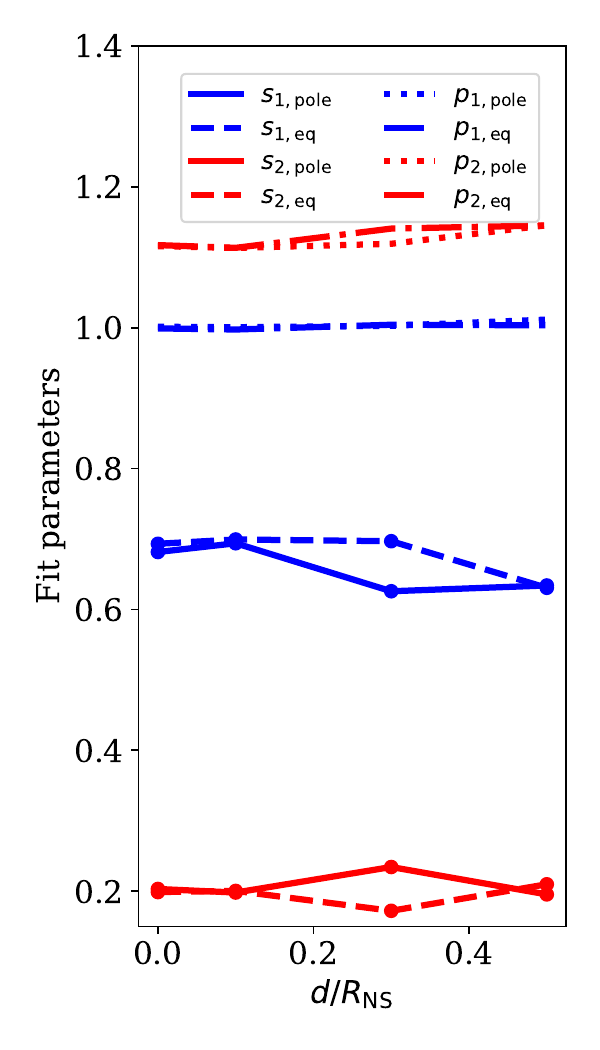}
    \end{minipage}    
    \caption{Evolution of hot area for different $d$-values at 50~kyr for magnetic configurations of BP models (left panel) and BE (right panel) with magnetised atmosphere \protect\cite{Potekhin2015SSRv} computed for $5\times 10^5$ photons.  }
    \label{fig:hot_areas_evol_pot}
\end{figure}

We further produce spectra for BP and BE models for two orientations and fit them with the sum of two black-body models. The results are shown in Figure~\ref{fig:hot_areas_evol_pot}. In this case it is necessary to receive $5\times 10^5$ photons to reliably resolve two black-body components. The temperature difference between the components is larger than in the non-magnetised case. The fraction between surface areas of two hot regions is three in this case. For the case of equatorial shift the spectral parameters are not very sensitive to the dipolar shift, and there is a small systematic trend of temperature with dipole shift $d$. In the case of the polar shift temperatures are slowly converging when $d$ is increased. It is opposite in the case of equatorial shift: temperatures are diverging with the dipole shift.

\subsection{Observational implications}
\label{s:observational_implications}

The results presented here are relevant primarily for middle-aged radio pulsars (ages $10^5$--$10^7$ years) and X-ray dim isolated neutron stars. X-ray observations (see e.g. \citealt{Zavlin2004ApJ,Harding2013FrPhy,Mereghetti2016ApJ,Rigoselli2018AA,Mereghetti2021ApJ,Rigoselli2022MNRAS}) showed that X-ray spectra of middle-aged radio pulsars are typically composed of power-law and one or two black-body components. The power-law corresponds to non-thermal emission produced by particle acceleration in the pulsar magnetosphere. The thermal emission is usually associated to bulk thermal emission (large emission area) and polar cap emission (small area typically corresponding to radius of $\approx 1$~km). Future detailed studies might discover more black-body components in the softest part of emission (below 1~keV). These could be compared with results obtained in this article to probe off-centred dipoles formed for example in single hemisphere dynamos. 

In the recent article by \cite{Rigoselli2022MNRAS,Yoneyama2019PASJ,Yoneyama2019AN} it was shown that typical $T_\mathrm{hot}/T_\mathrm{cold}\approx 2$ and $R_\mathrm{hot} / R_\mathrm{cold} \approx 0.1$. Similarly to their work we cannot reproduce these parameters. In the case of model B we obtain typical values of $T_\mathrm{hot}/T_\mathrm{cold}\approx 1.1$ and $R_\mathrm{hot} / R_\mathrm{cold} \approx 2$. Careful modelling of the NS atmosphere and magnetosphere will probably be required to produce values found in observations.

As for the XDINS, their spectra are much easier to interpret because they typically lacks non-thermal components and similarly seem to lack polar cap emission. In one particular case, RX J1856.5-3754 \cite{Sartore2012AA} the emission areas have comparable sizes which differ by a factor of four. This object was already suggested as a primary test for long-term magneto-thermal evolution \cite{PopovTaverna2017MNRAS, Popov2017JPhCS}. Initial simplified modelling \cite{PopovTaverna2017MNRAS} seemed to contradict the observed picture, while more detailed modelling \cite{deGrandis2021ApJ} agreed better with the observations. 

Another very interesting implication could be related to how radio pulsars age. If an NS is born with off-centred magnetic field, it means that the curvature of open field lines is significantly smaller at one pole. The dipole shift decreases with time and, thus, the curvature radius of open magnetic field lines increases. Small curvature radius of open field magnetic lines is favourable for pair production \citep{Timokhin2015ApJ} and operation of radio pulsar mechanism. NSs with increasing curvature radius of open field lines stop emitting as radio pulsar at timescales of $\propto 10^7$~years. Influence of curvature of open field lines on properties of radio pulsar emission was discussed by e.g. \cite{Szary2015MNRAS,Igoshev2016MNRAS}.

\section{Conclusions}
\label{s:conclusions}

For the first time we systematically develop a technique to write the initial conditions for off-centred dipoles, and studied evolution of such configurations using a code for magneto-thermal evolution of neutron stars. In this paper we concentrate only on crust-confined magnetic field configurations, assuming that there are no electric currents in the NS core. This assumption is partially supported by comparison of NS thermal evolution with detailed magneto-thermal simulations, see e.g. \cite{Dehman2023MNRAS}. In the future configurations where the magnetic field is also present in the core should be explored.

We found the following magneto-thermal evolution: 
\begin{itemize}
    \item We introduce two estimates to measure the dipole shift using the low-order spherical harmonics such as $d_{21}$ using dipole and quadrupole and $d_{31}$ using dipole and octupole. 
    \item Off-centred dipole configuration with no toroidal magnetic field evolve towards centred dipoles. The evolution timescale is inversely proportional to magnetic field strength: configurations with stronger field ($B_\mathrm{d} = 1.00\times 10^{14}$~G; C-models) evolve much faster than configurations with weak field ($B_\mathrm{d} = 1.00\times 10^{12}$~G; A-model). The dipole shift estimated using $d_{31}$ decays faster than the one estimated by $d_{21}$. The decay proceeds on Hall timescale $t_\mathrm{Hall} = 4\pi e n_0 L^2 / (cB)$ which translates into $t_d = 19\; \mathrm{Myr} \; (1.00 \times 10^{12}) / B$. 
    \item We also simulate the evolution of configurations with toroidal magnetic field. These configurations have initial poloidal magnetic field represented as a centred dipole and a strong dipolar crust confined toroidal component B0-0-tor and C0-0-tor. In these simulations the dipole shift initially grows with time. It reaches its maximum approximately at times when dipole shift start declining in corresponding simulations with no toroidal magnetic field B3-P and C3-P. Both estimates $d_{21}$ and $d_{31}$ initially grow simultaneously. Long term, the simulations with presence of toroidal magnetic field stays as an off-centred dipole significantly longer than configurations with no toroidal magnetic field.  
    \item The surface thermal maps are very different for configurations with weak magnetic field (below a few $10^{13}$~G) and stronger magnetic fields. In the case of weak magnetic field, they make the crust less transparent for thermal flux from the core around the magnetic equator. In these cases (A,B-models) the equator is cold. In the case of strong magnetic fields (C-models) the equator is heated due to release of magnetic energy via the Ohmic decay of electric currents. The exact boundary between strong and weak magnetic fields depends on crust conductivity and properties of neutrino cooling. This effect will be investigated in detail in the future. It will potentially help to better constrain the internal crust properties and magnetic field strengths.
\end{itemize}
The X-ray thermal spectra of NSs with weak and strong magnetic fields are very different. Assuming that we receive only $5\times 10^5$~photons (typical long exposure with modern X-ray instrumentation) in energy range 0.2-5~keV we find:
\begin{itemize}
    \item  Spectra of NS with weak magnetic fields are well-described with single BB independently of the dipole shift. The reason for this is that cold equator emits a few photons at energies where the photons are abundant. The temperature of NS bulk above and below the cold equator are basically not affected by the dipole shift. The total hot areas stays approximately the same. To probe the dipole shift in these cases would require collecting up to $5\times 10^6$ photons, and temperatures of two blackbody components are expected to be very similar.    
    \item  Spectra of NS with strong magnetic fields are typically described significantly better using 2BB model. The hot equator emits harder X-ray photons and contributes to the part of X-ray spectra where the bulk NS photons becomes rare (at 1-2~keV). The equator temperature is sensitive to the dipole shift and stronger dipole shift leads to increased number of harder X-ray photons. While fitting the 2BB model, we found that typical $p_1 / p_2 \approx 1.25$~--~$1.44$ and $p_1$ is sensitive to the dipole shift with higher $p_1$ indicating larger dipole shift for any rotational orientation of NS.
\end{itemize}
We model the soft X-ray lightcurves for all configurations and compute the maximum pulsed fractions. Analysing these results we found:
\begin{itemize}
    \item Larger pulsed fractions are seen for larger dipole shifts. The pulsed fraction becomes significantly larger (tens of percent in comparison to a few percents) when we consider stronger magnetic fields with hot equator. Even in this case, the larger dipole shift corresponds to larger pulse fraction. Shift of dipole in equatorial direction causes a very small pulsed fraction (fraction of percent). 
    \item For majority of our simulations the lightcurves are symmetric around the maxima. The only exception are models where the axial symmetry of thermal maps is significantly broken: B3-E, B5-E, B3-D1 models. 
    \item Lightcurves produced by C-models with strong magnetic fields are sinusoidal-like because the surface thermal map preserve significant degree of axial symmetry even in the case of C0-0-tor when separate smaller hot spots are developed as a result of the Hall instability.
\end{itemize}

Overall, we conclude that magnetic field strength (including the hidden toroidal magnetic field) is much easier to probe in X-ray observations of thermally emitting NS than the magnetic field topology e.g. strength and direction of dipole shift.

\section*{Acknowledgements}
API thanks Prof.\ Dmitry Yakovlev for answering multiple questions, Dr.\ A. Frantsuzova for enlightening discussions about statistics and anonymous referee for their insightful comments.
This work was supported by STFC grant no.\ ST/W000873/1. The authors would like to thank the Isaac Newton Institute for Mathematical Sciences for support and hospitality during the programme DYT2 ``Frontiers in dynamo theory: from the Earth to the stars" where work on this paper was undertaken. This programme was supported by EPSRC Grant Number EP/R014604/1. RH’s visit to the Newton Institute was supported by a grant from the Heilbronn Institute.

This work was performed using the DiRAC Data Intensive service at Leicester, operated by the University of Leicester IT Services, which forms part of the STFC DiRAC HPC Facility (www.dirac.ac.uk). The equipment was funded by BEIS capital funding via STFC capital grants ST/K000373/1 and ST/R002363/1 and STFC DiRAC Operations grant ST/R001014/1. DiRAC is part of the National e-Infrastructure.

\section*{Data Availability}

Code \texttt{Magpies} to model spectra and lightcurves of neutron stars is made publicly available at \url{https://github.com/ignotur/magpies}. Data used in this article will be provided under reasonable request by the corresponding author.
 



\bibliographystyle{mnras}
\bibliography{example} 




\appendix

\section{Initial condition for off-centre dipole}
\label{a:initial conditions}

We prepare the initial condition for \texttt{PARODY} code as a multiplication of two independent components: (1) an angular part and (2) a radial part. The angular part represents the external surface magnetic field matching an off-centred dipolar magnetic field in vacuum. The radial part satisfies boundary conditions at the crust-core interface and radial part of the boundary condition at the surface. 

The \texttt{PARODY} code works with the poloidal-toroidal decomposition, making it necessary to write an initial condition using the poloidal potential $\beta_p$. It is thus natural to present $\beta^{lm}_p$ as multiplication of a constant which describes the angular part and a function of radial coordinate $r$ which satisfies correct boundary conditions.

In the case of axisymmetric magnetic field configurations, elements $\beta^{lm}_p$ became zero for $m\neq 0$. The spherical harmonics are simplified to the Legendre polynomials $P_\ell (\cos \theta)$.

The expansion coefficients $b_{lm}^r$ for the radial component of the magnetic field $B_r$ are directly related to the expansion coefficients for poloidal potential $\beta_p^{lm}$:
\begin{equation}
b_{lm}^r = \frac{\ell(\ell+1)}{r} \beta^{lm}_p.
\label{eq:bl_br}
\end{equation}
Thus, if we expand the radial component into Legendre series (or spherical harmonics), we can transform expansion coefficients and derive the poloidal potential.

\subsection{Angular part}

The magnetic field outside the NS can be described as a magnetic dipole shifted by the vector $\vec d$ from the origin. In vector form it is described as the following \citep{Petri2016MNRAS}:
\begin{equation}
\vec B = \frac{B_\mathrm{eq} R_\mathrm{NS}^3}{|\vec R - \vec d|^3} \left[\frac{3\vec m \cdot (\vec R - \vec d) }{|\vec r - \vec d|^2} (\vec R - \vec d) - \vec m\right],
\end{equation}
where $\vec m$ is a unit vector in the direction of the magnetic moment and $B_\mathrm{eq}$ is the strength of the dipolar magnetic field at the equator. 

Alternatively, \cite{Petri2016MNRAS} used scalar magnetic field potential $\Psi$ which is related to the field as $\vec B = \vec \nabla \Psi$. Their potential for dipolar magnetic field is:
\begin{equation}
\Psi = \frac{\cos (\theta)}{r^2}.
\label{eq:psi}
\end{equation}
This is a truly scalar field and it is possible to shift the coordinate system to emulate the off-centred dipole. We consider three cases: (1) we shift the dipole along the dipole axis by value $d$, (2) we shift dipole along the magnetic equator, and (3) we shift dipole along the diagonal axis.

\subsubsection{Shift along the north-south direction}

\begin{figure}
    \centering
    \includegraphics[width=\columnwidth]{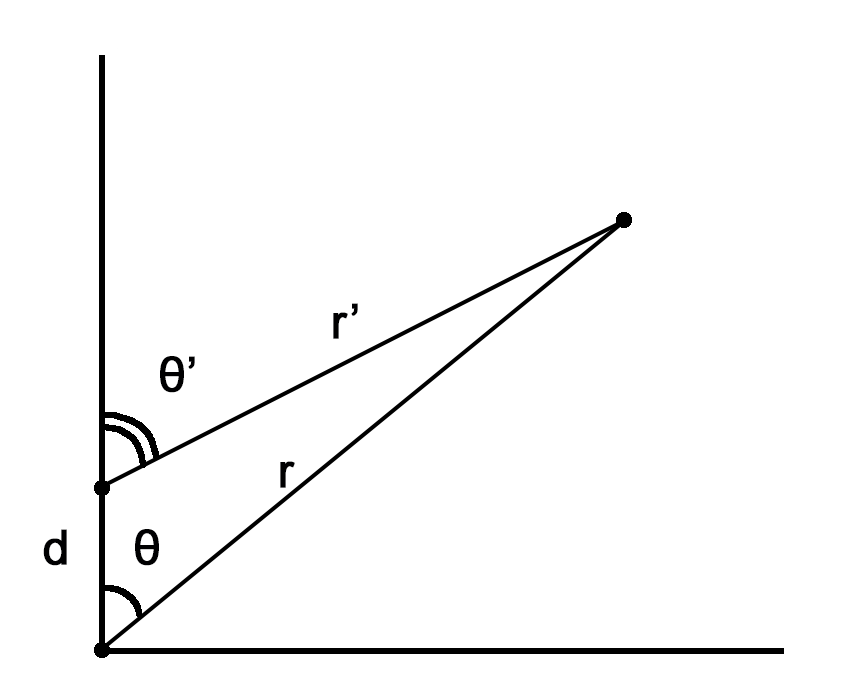}
    \caption{Illustration of angles involved in off-centred dipole problem when dipole is shifted along north-south direction.  }
    \label{fig:theta_r}
\end{figure}

In order to find numerical coefficients for the expansion we write $\Psi$ potential for the off-centred dipole using new coordinates $r_1$ and $\theta_1$. We demonstrate a relation between different angles in Figure~\ref{fig:theta_r}. These coordinates are related to the original spherical coordinate system with values $r$ and $\theta$ according to:
\begin{equation}
r_1 = \sqrt{r^2 + d^2 - 2 r d \cos(\theta)}.
\end{equation}
Functions of angle $\theta_1$ are computed as follows:
\begin{equation}
\cos \theta_1  =  (r\cos(\theta) - d)  /   r_1.
\end{equation}

In this new coordinate system the components of magnetic field can be computed as:
\begin{equation}
B_r  = \frac{\cos \theta_1 \left(3 d \cos{\theta } - 3 r\right)}{r_1^5} + \frac{\cos{\left(\theta \right)}}{r_1^3},
\end{equation}
\begin{equation}
B_\theta = - \frac{3 d \cos \theta_1 \sin{\theta }}{r_1^4} - \frac{ \sin{\left(\theta \right)}}{r_1^3}.
\end{equation}

\begin{figure*}
    \centering
    \begin{minipage}{0.49\linewidth}
    \includegraphics[width=\columnwidth]{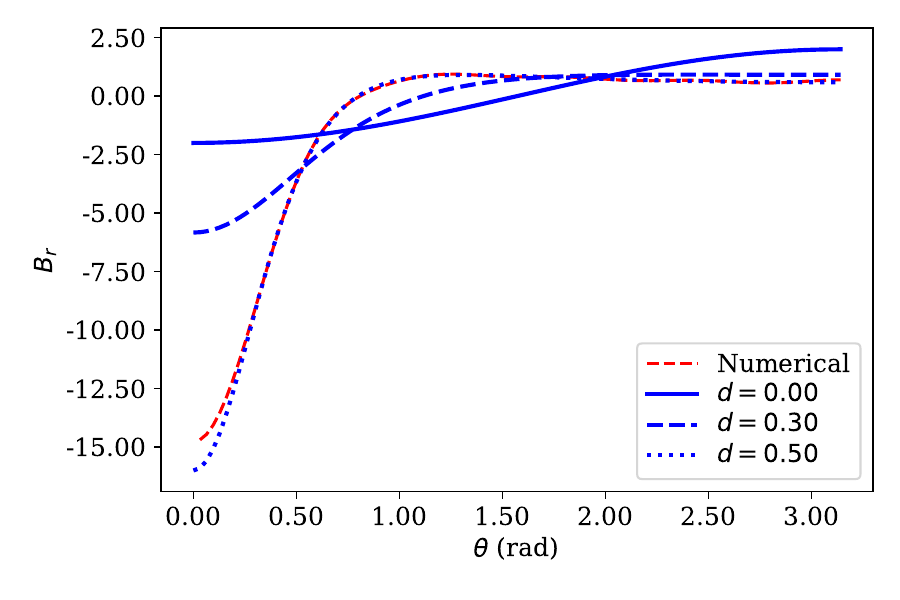}
    \end{minipage}
    \begin{minipage}{0.49\linewidth}
    \includegraphics[width=\columnwidth]{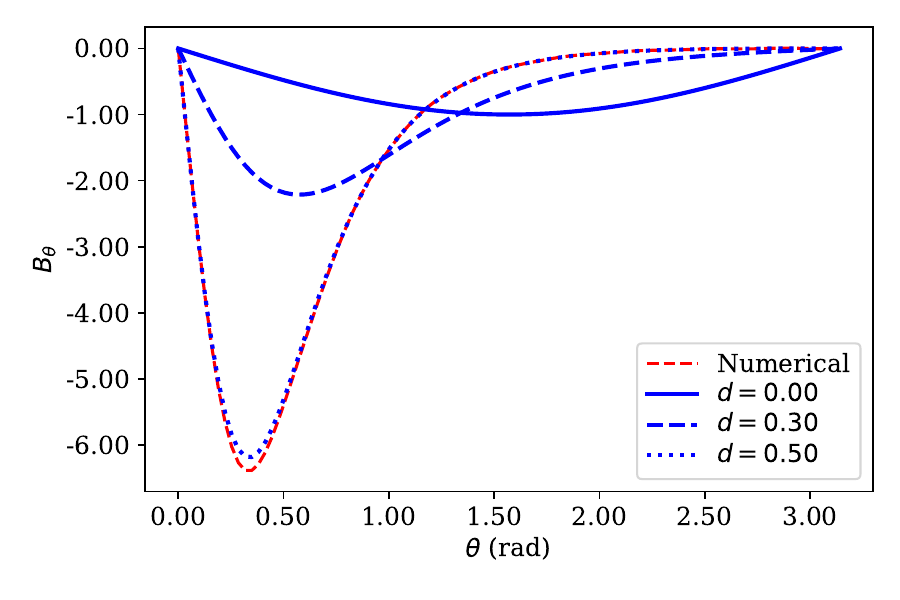}
    \end{minipage}
    \caption{Components $B_r$ (left panel) and $B_\theta$ of the surface magnetic field for off-centred dipole when dipole is shifted along the north-south direction. The red dashed line shows the values computed by numerical differentiation of the $\beta_p$ potential expanded as a Legendre series with $N_l = 10$ for $d=0.5$. }
    \label{fig:surface_field}
\end{figure*}

We show $B_r$ and $B_\theta$ components at the NS surface for different shifts $d$ in Figure~\ref{fig:surface_field}.
Because it is an axisymmetric configuration, we expand the radial part of the magnetic field at the neutron star surface into a Legendre polynomial series:
\begin{equation}
B_r = \sum_{l=1}^\infty b_l^r P_\ell (\cos \theta).
\end{equation}
Here we assume that $R_\mathrm{NS} = 1$. We find the numerical values of $b_l^r$ using the orthogonality property of Legendre polynomials: 
\begin{equation}
b^r_l (R_\mathrm{NS}) = \frac{2\ell + 1}{2}\int_0^\pi B_r (r_\mathrm{NS}, \theta) P_\ell (\cos\theta) \sin \theta d\theta.
\end{equation}
The numerical coefficient in front of the integration appears because $\int P_\ell (\cos\theta)^2 \sin \theta d\theta = 2 / (2\ell+1)$.
In \texttt{PARODY} the spherical harmonics are normalised using numerical coefficients \citep{dormy1997}:
\begin{equation}
C_\ell^m = \sqrt{(2-\delta_{m0}) (2\ell + 1) \frac{(\ell-m)!}{(\ell+m)!}},
\label{eq:clm_parody}
\end{equation}
which translates to:
\begin{equation}
C_\ell = \sqrt{2\ell + 1}   
\label{eq:cl}
\end{equation}
for axially symmetric configurations. We summarise the numerical coefficients $b_{l0}^r$ for different values of $d$ in Table~\ref{t:coeff}. These coefficients simply follow a relation:
\begin{equation}
b^r_{\ell 0} = d^{({\ell-1})}.
\end{equation}

We relate these numerical coefficients to the expansion of poloidal potential using eq. (\ref{eq:bl_br}) and taking into account the normalisation eq. (\ref{eq:cl}):
\begin{equation}
\beta_p^{\ell 0} = \frac{b_\ell^r}{\ell(\ell+1)\sqrt{2\ell+1}},
\label{blr_convr}
\end{equation}

We show how our numerical potential computed with these coefficients translates into $B_r$ and $B_\theta$ in Figure~\ref{fig:surface_field}.

\begin{table}
    \caption{Coefficients of expansion for dipole shifted in the north-south direction.}
    \label{t:coeff}
    \centering
    \begin{tabular}{ccccc}
    \hline
    $l$  & $d=0$ & $d=0.1$ & $d=0.3$ & $d=0.5$ \\
    \hline
     1   & -1.00 & -1.000 & -1.000  & -1.000 \\
     2   & 0.000  & -0.100 & -0.300  & -0.500 \\
     3   & 0.000  & -0.010 & -0.090  & -0.250  \\
     4   & 0.000  & -0.001 & -0.027  & -0.125 \\
     5   & 0.000  & 0.000    & -0.008  & -0.062 \\
     6   & 0.000  & 0.000    & -0.001  & -0.031 \\
     7   & 0.000  & 0.000    & 0.000     & -0.015 \\
     8   & 0.000  & 0.000    & 0.000     & -0.008 \\
     9   & 0.000  & 0.000    & 0.000     & -0.004 \\
     10  & 0.000  & 0.000    & 0.000     & -0.002 \\
     \hline
    \end{tabular}
\end{table}

\subsubsection{Shift along magnetic equator}

Here we introduce a spherical coordinate system which is related to the Cartesian coordinates as:
\begin{equation}
\begin{array}{ccc}
x & = & r \sin\theta \sin \phi, \\
y & = & r \sin\theta \cos \phi, \\
z & = & r \cos \theta. \\
\end{array}
\end{equation}
We rewrite the potential eq. (\ref{eq:psi}) for the Cartesian coordinate system as:
\begin{equation}
\Psi = \frac{z}{(x^2 + y^2 + z^2)^{3/2}}.
\end{equation}
A potential in this form can be easily modified to take into account a shift in $x$ or $y$ direction as follows:
\begin{equation}
\Psi = \frac{z}{\left[(x-d)^2 + y^2 + z^2\right]^{3/2}}.
\end{equation}
Now we find derivatives with respect to $x, y$ and $z$ and use the chain rule:
\begin{equation}
B_r = \frac{\partial \Psi}{\partial x} \frac{\partial x}{\partial r} + \frac{\partial \Psi}{\partial y} \frac{\partial y}{\partial r} + \frac{\partial \Psi}{\partial z} \frac{\partial z}{\partial r}.
\label{s:eq_Br_pot}
\end{equation}
If we introduce an auxiliary variable $r_2$
\begin{equation}
r_2 = \sqrt{  (x-d)^2 + y^2 + z^2 },
\label{eq:r2}
\end{equation}
the eq. (\ref{eq:r2}) becomes:
$$
B_r = \frac{3 z \left(d - x\right) \sin \theta \sin \phi}{r_2^{5}} - \frac{3 y z \sin \theta \cos \phi}{r_2^{5}} \hspace{1cm}
$$
\begin{equation}
 {\ } - \frac{3 z^{2} \cos\theta}{r_2^{5}} + \frac{\cos\theta}{r_2^{3}}.
\end{equation}
We compute this radial magnetic field at the top of the NS crust, i.e.\ at the surface of a sphere with unit radius. After this, we expand this radial magnetic field at the NS surface using the spherical harmonics:
\begin{equation}
B_r = \sum_{l=1}^{N_l} \sum_{m=-l}^{l} b^r_{lm} Y_{lm} (\theta, \phi).
\end{equation}
In \texttt{Python} package \texttt{scipy} the spherical harmonics are normalised as:
\begin{equation}
C_\ell^m = \sqrt{\frac{2\ell+1}{4\pi}\frac{(\ell-m)!}{(\ell+m)!}}    
\end{equation}
which is different from \texttt{PARODY} eq. (\ref{eq:clm_parody}). Thus we need to take into account this difference while performing the conversion.
We convert $b^r_{lm}$ into $b_{lm}$ using:
\begin{equation}
\beta_p^{l0} = \frac{b_{l0}^r}{\ell(\ell+1)\sqrt{4\pi}}    
\label{eq:betabrm0}
\end{equation}
and for $m\neq 0$ we obtain:
\begin{equation}
\beta_p^{lm} = \frac{2b_{lm}^r}{\ell(\ell+1)\sqrt{8\pi}}    
\label{eq:betabrm1}
\end{equation}
Additional factor of 2 in the nominator appears because while $\texttt{scipy}$ works with positive and negative m, \texttt{PARODY} only uses $0\leq m\leq l$. Therefore, $b_{lm}^r$ should contribute twice.

\subsubsection{Shift in diagonal direction}

For the diagonal shift we write the potential $\Psi$ as:
\begin{equation}
\Psi = \frac{z - d/\sqrt{2}}{\left[(x - d/\sqrt{2})^2 + y^2 + z^2\right]^{3/2}}.
\end{equation}
We can then evaluate the derivatives $\partial \Psi / \partial x$, $\partial \Psi / \partial y$ and $\partial \Psi / \partial z$. 
We write the surface components of radial magnetic field using eq. (\ref{s:eq_Br_pot}). After this for each value of the shift $d$ we compute the strength of radial magnetic field at the NS surface using a uniform grid in $\theta$ and $\phi$. In order to estimate the coefficients of the expansion we numerically find integrals:
\begin{equation}
b_{lm}^r = \int_0^\pi \int_0^{2\pi} B_r (\theta, \phi)\; Y_{lm}^* (\theta, \phi) \sin \theta d\theta d\phi.
\end{equation}
We convert these coefficients for expansion of the surface magnetic field into coefficients for potential using eqs. (\ref{eq:betabrm0}) and (\ref{eq:betabrm1}).

\subsection{Radial part of initial condition}

For the radial part, we consider the potential in the form of polynomials which satisfy correct boundary and additional conditions. We set the following  conditions:
\begin{equation}
\begin{array}{ccc}
  \beta_p^{lm} (1)   & = & \beta_p^{lm}, \\
  \beta_p^{lm} (r_c) & = & 0, \\
  \frac{\partial \beta_p^{lm}}{\partial r} (1) + \frac{(\ell+1)}{r} \beta_p^{lm} (1) &= & 0,
\end{array} \label{eq:cond}   
\end{equation}
where $r_c$ is the core-crust boundary and $\beta_p^{lm}$ is a value found in the previous sections.

We are looking for polynomials of the form:
\begin{equation}
\beta_p^{lm} = \frac{(a + b r + c r^2)}{r}.
\end{equation}
Overall, eq. (\ref{eq:cond}) written for our desired polynomial solution becomes:
\begin{equation}
\begin{array}{ccc}
a + b + c & = & \beta_p^{lm},  \\
a + b r_c + c r_c^2 & = & 0, \\
a l + (l+1) b + (l+2) c & = & 0.\\
\end{array}
\end{equation}
This system of linear equations can be solved analytically to yield:
\begin{equation}
a = \frac{b_{l} \ell r_{c}^{2} - b_{l} \ell r_{c} + b_{l} r_{c}^{2} - 2 b_{l} r_{c}}{r_{c}^{2} - 2 r_{c} + 1},
\end{equation}
\begin{equation}
b = \frac{- b_{l} \ell r_{c}^{2} + b_{l} \ell + 2 b_{l}}{r_{c}^{2} - 2 r_{c} + 1},
\end{equation}
\begin{equation}
c = \frac{b_{l} \ell r_{c} - b_{l} \ell - b_{l}}{r_{c}^{2} - 2 r_{c} + 1}.
\end{equation}

\section{Symmetries in the problem}
\label{A:symmtries}

It is obvious by inspection that the Hall term is not invariant under ${\vec B}\rightarrow-{\vec B}$. As demonstrated by \cite{Hollerbach2002MNRAS}, for axisymmetric solutions it turns out that the poloidal and toroidal components separate out in such a way that the poloidal component does satisfy $\pm \beta_p$ being equivalent, and it is only the toroidal component $\beta_t$ which violates this symmetry. We wish to show here that for fully three-dimensional solutions this is not the case, and the general evolution depends on the signs of both components.

We start with just the Hall term alone:
\begin{equation}
\frac{\partial \vec B}{\partial t} = - \nabla \times \left[f(r) (\nabla \times \vec B) \times \vec B \right].
\end{equation}
In fact, the radial profile $f(r)$ is irrelevant for this symmetry question we are addressing here, so we drop $f(r)$ in what follows.
Considering the poloidal-toroidal expansion, and writing equations for spherical harmonics we get two separate equations:
\begin{equation}
\sum_{l,m} \frac{l(l+1)}{r^2} \frac{\partial \beta_p^{lm}}{\partial t} = -\vec r \cdot  \nabla \times \left[ (\nabla \times \vec B) \times \vec B\right],
\label{eq:pol}
\end{equation}
\begin{equation}
\sum_{l,m} \frac{l(l+1)}{r^2} \frac{\partial \beta_t^{lm}}{\partial t} = -\vec r \cdot  \nabla \times  \nabla \times \left[ (\nabla \times \vec B) \times \vec B\right],
\label{eq:tor}
\end{equation}
where $\vec r$ is the unit vector in the radial direction.
We introduce the abbreviations $C = \nabla \times $, $g = \beta_t$ and $h = \beta_p$. Using these we can write the magnetic field as:
\begin{equation}
\vec B = C (g\vec r)  + C^2 (h\vec r),  
\end{equation}
from the basic definition of toroidal and poloidal components.
We can then write the Hall term as:
$$
(\nabla \times \vec B) \times \vec B = C^2 (g\vec r)\times C(g\vec r) +  C^2 (g\vec r)\times C^2(h\vec r)
$$
\begin{equation}
 + C^3 (h\vec r)\times C(g\vec r)  + C^3 (h\vec r)\times C^2(h\vec r).
\end{equation}
We schematically rewrite eqs. (\ref{eq:pol}, \ref{eq:tor}) as:
\begin{equation}
\frac{\partial h}{\partial t} = C (g,g) + C(g,h) + C(h,g) + C(h,h),    
\end{equation}
\begin{equation}
\frac{\partial g}{\partial t} = C^2 (g,g) + C^2 (g,h) + C^2(h,g) + C^2(h,h),
\end{equation}
where $C(g,g)$ stands for $C(C^2(g\vec r)\times C(g\vec r))\cdot{\vec r}$, etc.

Now, suppose it turned out that the terms $C(g,g)$ and $C(h,h)$ for $\partial h/\partial t$, and $C^2 (g,h)$ and $C^2(h, g)$ for $\partial g / \partial t$ were identically zero. That would then leave:
\begin{equation}
\frac{\partial h}{\partial t} =  C(g,h) + C(h,g),
\end{equation}
\begin{equation}
\frac{\partial g}{\partial t} = C^2 (g,g) + C^2(h,h).
\end{equation}
These equations are clearly invariant under $h\rightarrow -h$, but the sign of $g$ does matter. However, if any of the missing terms were not identically zero after all, then the $h\rightarrow -h$ symmetry is violated, and the sign of $h$ matters as well.

We wish to show then that these four terms are identically zero for axisymmetric solutions, but not for fully three-dimensional ones.
Consider for example the $C(g,g)$ term:
\begin{equation}
C(g,g) = C \left[C^2 (g\vec r) \times C (g\vec r) \right] \cdot \vec r.
\end{equation}
For axisymmetry we have $C(g\vec r) = (0,0,...)$ for the $r,\theta,\phi$ components of the vector. That is, the $r$ and $\theta$ components are identically zero, and only the $\phi$ component is non-zero. For the second curl, we similarly get $C^2 (g \vec r) = (...,...,0)$. The cross product therefore becomes:
\begin{equation}
C^2(g\vec r) \times C(g \vec r) = (..., ..., 0).
\label{curls}
\end{equation}
Recall then that for $C(g,g)$ we want one more curl of this whole combination, and then specifically the $r$ component of that curl. Well, if the $\phi$ component of (\ref{curls}) is zero, as we just showed, and if $\frac{\partial}{\partial\phi}\equiv0$ for axisymmetric solutions, we easily obtain $C(g,g)\equiv0$. The other three terms are similarly seen to be identically zero in the axisymmetric case. This reconfirms the symmetry results of \cite{Hollerbach2002MNRAS}, who used a much simpler form of poloidal-toroidal decomposition that only applies for axisymmetric solutions. However, we can also easily see at this point that if we no longer have $\frac{\partial}{\partial\phi}\equiv0$, then in general $C(g,g)$ will not be identically equal to zero. If we wanted we could similarly consider the other three terms as well, but having even just one of them no longer vanish identically is already enough to demonstrate that for fully three-dimensional solutions the sign of the poloidal field component also matters.

The relevance of these symmetry considerations to the off-centre dipoles computed here is as follows: if the dipole is displaced in the purely polar direction, the situation remains axisymmetric, so the dipole's sign doesn't matter, which is equivalent to saying that it does not matter whether it is shifted up or down. This is why we only computed positive displacements in this case. In contrast, if the dipole is displaced in a diagonal direction, the situation becomes non-axisymmetric, so at least potentially the dipole's sign does matter, so it is necessary to consider displacements in both directions.
 

\bsp	
\label{lastpage}
\end{document}